\documentclass[conference]{IEEEtran}
\usepackage[latin9]{inputenc}
\usepackage{fancyhdr}
\usepackage{xcolor}
\usepackage{pdfcolmk}
\usepackage{textcomp}
\usepackage{amsmath}
\usepackage{amssymb}
\usepackage{graphicx}
\PassOptionsToPackage{normalem}{ulem}
\usepackage{ulem}

\makeatletter

\providecolor{lyxadded}{rgb}{0,0,1}
\providecolor{lyxdeleted}{rgb}{1,0,0}

\DeclareRobustCommand{\lyxsout}[1]{\ifx\\#1\else\sout{#1}\fi}

\IEEEoverridecommandlockouts
\usepackage{cite}
\usepackage{amsfonts}\usepackage{algorithmic}
\usepackage{subfigure}
\usepackage{textcomp}
\usepackage{xcolor}
\def\BibTeX{{\rm B\kern-.05em{\sc i\kern-.025em b}\kern-.08em
    T\kern-.1667em\lower.7ex\hbox{E}\kern-.125emX}}
\usepackage{fancyhdr}
 
\DeclareTextSymbolDefault{\textquotedbl}{T1}

\IEEEoverridecommandlockouts

\makeatother

\begin{document}
\title{General Conditions to Realize Exceptional Points of Degeneracy in
Two Uniform Coupled Transmission Lines}
\author{\IEEEauthorblockN{Tarek Mealy and Filippo Capolino } \IEEEauthorblockA{\textit{Department of Electrical Engineering and Computer Science,
University of California, Irvine, CA 92697 USA} \\
 tmealy@uci.edu and f.capolino@uci.edu}}
\maketitle
\thispagestyle{fancy}
\begin{abstract}
We present the general conditions to realize a fourth order exceptional
point of degeneracy (EPD) in two \textit{uniform} (i.e., invariant
along $z$) lossless and gainless coupled transmission lines (CTLs),
namely, a degenerate band edge (DBE). Until now the DBE has been shown
only in periodic structures. In contrast, the CTLs considered here
are uniform and subdivided into four cases where the two TLs support
combinations of forward propagation, backward propagation and evanescent
modes (when neglecting the mutual coupling). We demonstrate for the
first time that a DBE is supported in \textit{uniform} CTLs when there
is proper coupling between: (i) propagating modes and evanescent modes,
(ii) forward and backward propagating modes, or (iii) four evanescent
modes (two in each direction). We also show that the loaded quality
factor of \textit{uniform} CTLs exhibiting a fourth order EPD at $k=0$
is robust to series losses due to the fact that the degenerate modes
do not advance in phase. We also provide a microstrip possible implementation
of a uniform CTL exhibiting a DBE using periodic series capacitors
with very sub-wavelength unit-cell length. Finally, we show an experimental
verification of the existence DBE for a microstrip implementation
of a CTL supporting coupled propagating and evanescent modes. 
\end{abstract}


{\let\thefootnote\relax\footnotetext{This material is based on work supported by the Air Force Office of Scientific Research award number FA9550-15-1-0280 and FA9550-18-1-0355, and by the National Science Foundation under award NSF ECCS-1711975. The authors are thankful to DS SIMULIA for providing CST Studio Suite that was instrumental in this study}}

\section{Introduction}

\IEEEPARstart{E}{xceptional} points of degeneracy (EPDs) are points
in parameters space where two or more eigenmodes of a waveguide coalesce
into a single eigenmode. The dispersion relation of eigenmodes in
a waveguide that exhibits an EPD with order $m$, where $m$ is the
number of coalescing eigenmodes, has the behavior of $(\omega-\omega_{e})\propto(k-k_{e})^{m}$
near the EPD at $(\omega_{e},k_{e})$ \cite{F3}-\cite{2017theory}.
Here $\omega$ and $k$ are the angular frequency and the wavenumber,
respectively, and the EPD is denoted by the subscript $e$. Such dispersion
behavior is accompanied by a severe reduction in the group velocity
of waves propagating in those structures and a tremendous increase
in local density of states \cite{othman2016giant} resulting in a
giant increase in the loaded quality factor of the structure \cite{F2}-\cite{nada2018giant}.
Indeed for a lossless waveguide exhibiting an EPD of order order $m$
not only the group velocity $v_{g}=\partial\omega/\partial k$ vanishes,
but all of its derivatives $\partial v_{g}^{i}/\partial k^{i}$ with
$i<m-1$ vanish as well \cite{figotin2003oblique}.

In general, EPDs occur in coupled resonator systems and in coupled-multimode
waveguides. Recently the occurrence of EPDs has been shown in a \textit{single}
resonator where one of its elements is time modulated \cite{Hamid_EPD_LTV}.
In this paper we focus on EPDs occurring in multimode waveguides.
Furthermore, there are a few types of EPDs, some involve the simultaneous
presence of loss and gain, like in PT-symmetric systems \cite{el2007theory}\nocite{ruter2010observation}-\cite{othman2016theory}.
Here however we focus on EPDs that do not require loss and gain to
occur, namely we focus on the regular band edge and on the degenerate
band edge (DBE), that is a fourth order EPD introduced a few years
ago by Figotin and Vitebskiy in layered anisotropic crystals \cite{F3},
\cite{F2}.

Recent work has shown that the DBE can be engineered in various types
of \textit{periodic} guiding systems. The DBE is a fourth order EPD
existing in periodic waveguides without loss and gain. It has been
shown to exist in photonic crystals \cite{F3}, \cite{F4}, \cite{F5},
circular waveguides with periodic inclusions \cite{othman2015demonstration},
two coupled substrate integrated waveguides \cite{zheng2019design},
two coupled periodic transmission lines \cite{F6}-\cite{F7}, ladder
circuits \cite{sloan2017theory}, and integrated coupled optical waveguides
\cite{2017theory,burr2013degenerate}. The first experimental demonstration
of the existance of the DBE in periodic waveguides at radio frequency
was shown in \cite{othman2017experimental}, and recently extended
to periodic coupled microstrips \cite{abdelshafy2018exceptional}.
Structures exhibiting DBEs have been proposed recently for a wide
range of applications such as, for example, high quality factors photonic
crystals \cite{nada2018giant}, high power electron-beam devices \cite{yazdi2017new}-\cite{abdelshafy2018electron},
RF oscillators \cite{oshmarin2016oscillator} and lasers \cite{veysi2017theory}.

There are only a few ways to obtain EPDs in \textit{\textcolor{black}{uniform}}
waveguides. The simplest second order EPD is found in uniform waveguides
at the modal cutoff frequency where two modes, the forward and backward
modes, coalesce at $k=0$, forming an EPD of order 2 that is called
``regular'' band edge \cite{mealy2019degeneracy}. Another way to
realize second order EPDs in \textit{uniform} coupled transmission
lines (CTLs) is based on parity-time (PT-) symmetry \cite{el2007theory},
\cite{bender1998real} which implies using a balanced and symmetrical
distribution of gain and loss \cite{othman2016theory}. In contrast
to these two types of second order EPD, in this paper we show there
are other ways to realize EPDs of fourth order in two lossless/gainless
\textit{uniform} CTLs at $k=0$. Therefore this paper shows for the
first time how to realize a DBE at $k=0$ in uniform transmission
lines (Fig. 1) since previously the DBE was shown only in periodic
waveguides\cite{F3,2017theory,othman2015demonstration,burr2013degenerate,abdelshafy2018exceptional}.
This paper also shows how to locate a regular band edge (an EPD of
order 2) at any $k$, in \textit{\textcolor{black}{uniform}} waveguides
(Fig. 1).

In Section II, we discuss briefly all possible EPDs that may exist
in two \textit{uniform} CTLs, and their general necessary and sufficient
conditions. In Section III, we show the necessary and sufficient conditions
to realize fourth order EPD in two uniform, lossless, CTLs in term
of their per-unit-length parameters and we show all possible typologies
that may support a fourth order EPD, namely a DBE, at $k=0$. We also
show that CTLs of finite length make formidable resonators that exhibit
an $L^{5}$ scaling of the quality factor with the CTL length $L$.
Finally we show the effect of CTL losses on the occurrence of the
DBE and on the quality factor and show that series losses affect the
DBE much less than shunt losses. In Section IV, we present an example
of\textit{ uniform} CTLs that support a DBE at $k=0$ and we also
provide a microstrip possible implementation of such uniform CTLs
exhibiting the DBE using a series per-unit-length inductance realized
with a very sub-wavelength unit-cell length. In Section V we show
two experimental validations of the occurrence of the DBE in uniform
CTLs, using periodic capacitive loading with subwavelength period,
approximating (in a metamaterials sense) the uniform CTL. The findings
in this paper open up new ways to conceive distributed oscillators,
leaky wave antennas, and radiating leaky wave antennas with extreme
tunability, waveguide-based sensors, etc.

\section{System Description of Uniform Coupled Waveguides}

\begin{figure}
\centering \subfigure[]{\includegraphics[width=0.225\textwidth]{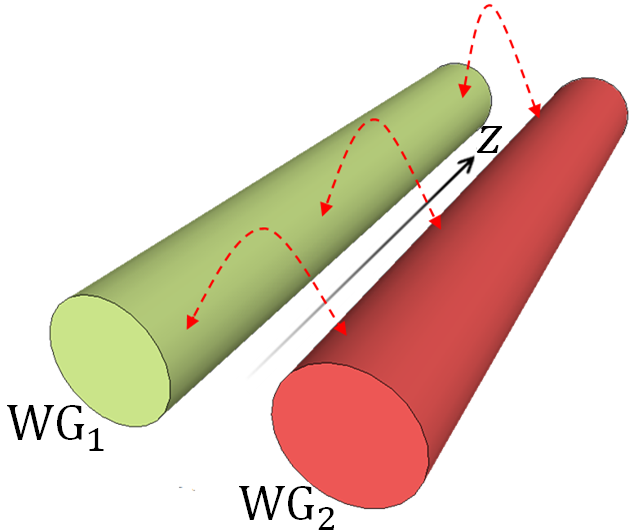}\label{fig:CTL_Uni}}
\subfigure[]{\includegraphics[width=0.225\textwidth]{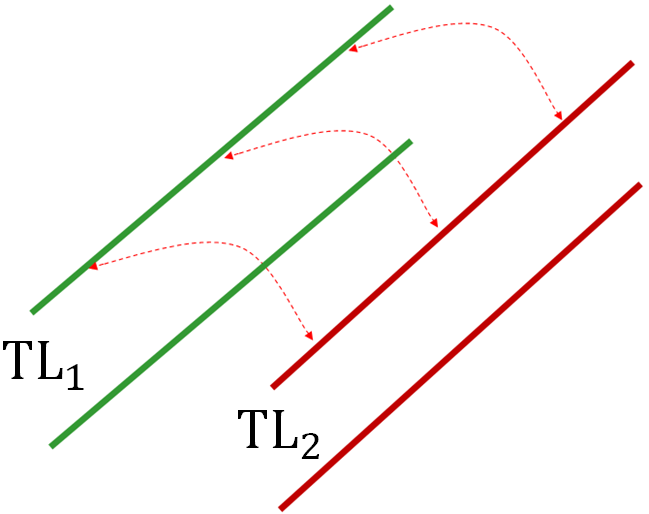}\label{fig:TL_Model}}

\centering \subfigure[]{\includegraphics[width=0.24\textwidth]{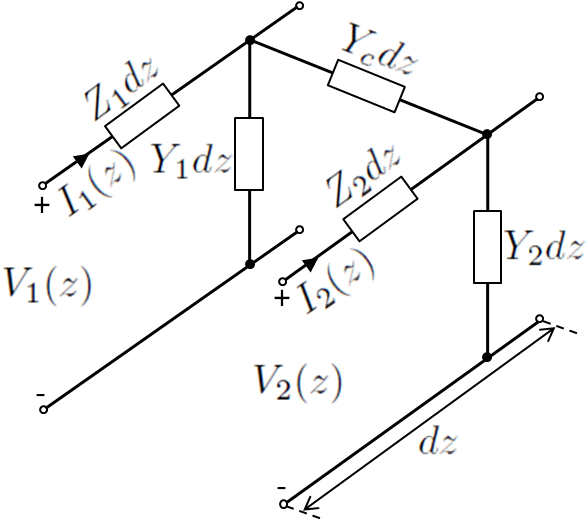}\label{fig:CCT_Model}}
\subfigure[]{\includegraphics[width=0.225\textwidth]{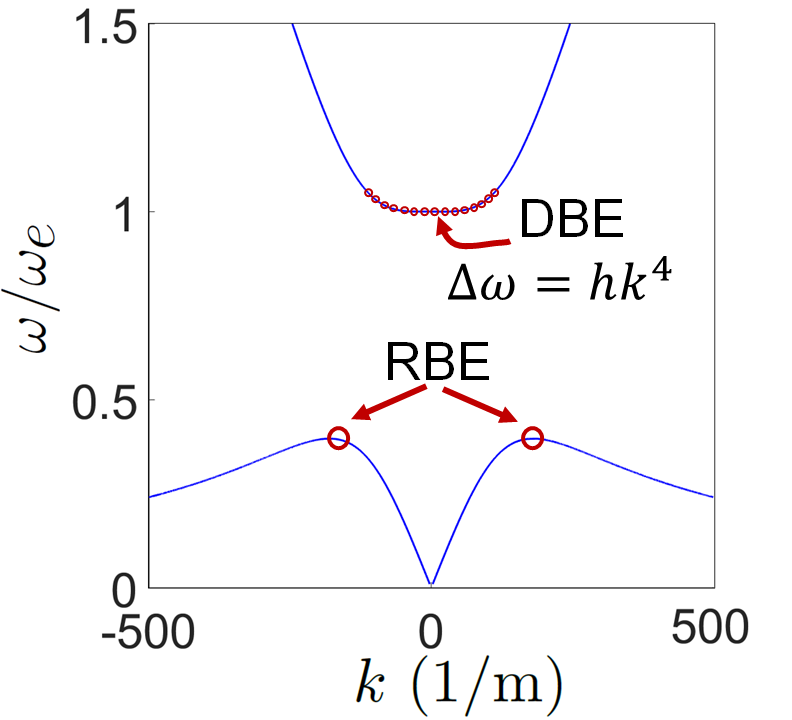}
\label{fig:CCT_Model-1}}
\raggedright{}\caption{(a) Two uniform coupled waveguides supporting four modes (two in each
direction). Modes have wavenumbers satisfying the $k$ and $-k$ symmetry,
due to reciprocity. (b) Equivalent coupled transmission line (CTL)
model describing the propagation of the four modes in the two uniform
coupled waveguides. (c) Generalized per-unit-length distributed equivalent
circuit model for the CTL. Coupling is represented by the distributed
(i.e., per-unit-length) admittance $Y_{c}$ . In this paper we determine
the necessary and sufficient conditions that the five reactances shall
satisfy for the CTLs to exhibit a DBE, i.e., a fourth order degeneracy.
(d) Representation of a dispersion diagram (showing only the branches
of purely-real wavenumber) reporting two important features: the DBE
at $k_{e}=0$ and $\omega=\omega_{e}$ (that is a fourth order EPD),
and a regular band edge at $\omega=0.4\omega_{e}$, with a non-vanishing
wavenumber of $k=\pm147.5\mathrm{m^{-1}}$ (a second order EPD).}
\end{figure}
Consider the two \textit{uniform} waveguides schematically shown in
Fig. \ref{fig:CTL_Uni}, where each waveguide (when uncoupled) supports
either a forward propagating mode, a backward propagating mode (where
group velocity and phase velocity have opposite sign) or an evanescent
mode; along each positive and negative $z$-direction due to reciprocity.

Let $V_{n}$ and $I_{n}$ , with $n=1,2$ , be the equivalent voltage
and current in each TL of Fig. \ref{fig:TL_Model}, describing the
spatial evolution of electromagnetic waves along the $z$-direction.
It is convenient to introduce the two-dimensional vectors $\mathbf{V}(z)=[\begin{array}{cc}
V_{1}(z)\ , & V_{2}(z)\end{array}]^{\mathrm{T}}$, $\mathbf{I}(z)=[\begin{array}{cc}
I_{1}(z)\ , & I_{2}(z)\end{array}]^{\mathrm{T}}$, where the superscript $\mathrm{T}$ represents the transpose operation. 

When the two transmission lines are not coupled they support four
independent modes that are described by four distinct wavenumbers
$k_{1}^{'}$, $k_{2}^{'}$ and $-k_{1}^{'},$ $-k_{2}^{'}$ and their
voltage and current are written as

\begin{equation}
\begin{array}{lc}
V_{n}(z)\propto e^{\pm jk_{n}^{'}z}, & I_{n}(z)\propto e^{\pm jk_{n}^{'}z},\end{array}\label{eq:Field_Exp-1-1}
\end{equation}
where, the modal wavenumbers $k_{n}^{'}$ , with $n=1,2$, are are
generally written as $k_{n}^{'}=\beta_{n}-j\alpha_{n}$, where $\beta_{n}$
and $\alpha_{n}$ are the phase propagation and attenuation constants,
respectively, and they determine the type of mode; for example, a
wavenumber $k^{'}$ that possesses only the imaginary part $\alpha$
is an evanescent mode. Forward modes are determined by $\beta\alpha>0$,
whereas ``backward'' propagating modes have $\beta\alpha<0$ (hence,
backward propagating modes have phase and group velocities with opposite
directions). 

The circuit \textit{equivalent} model for an infinites\textcolor{black}{imal-length
of a wavguide is represented by generic per-unit-length distributed
parameters as shown in Fig. \ref{fig:CCT_Model}. There, $Z_{1},$
$Z_{2},$ $Y_{1},$ $Y_{2}$ and $Y_{c}$ may be inductive or capacitive
impedances and admittances. In this paper, for the sake of brevity,
we do not consider magnetic induction coupling between the two TLs,
i.e., we only consider shunt per-unit-length inductive or capacitive
coupling $Y_{c}$ shown in Fig. \ref{fig:CCT_Model}. Coupling due
to magnetic induction between two nearby lines could be investigated
using the same mechanism and formulation used in this paper and it
is not treated here. It can be neglected in several cases, when the
separation between the two lines is very large, for examples, or for
the case studied in Sec. IV, where the coupling is due to the physical
connection between the $1^{st}$ and $2^{nd}$ $\mathrm{TL}$. }

\textcolor{black}{We assume that $Z_{1}$ and $Z_{2}$ may be either
capacitive or inductive impedances, as well as $Y_{1},$ $Y_{2}$
can be either capacitive or inductive, where the subscripts 1 and
2 are used to describe the parameters in the first and second transmission
line $\mathrm{TL_{1}}$ and $\mathrm{TL_{2}}$, respectively. We recall
that a single TL (say $\mathrm{TL_{1}}$ for example) supports backward
waves} if $Z_{1}$ is capacitive and $Y_{1}$ is inductive. Furthermore,
a TL (say $\mathrm{TL_{1}}$ for example) supports evanescent waves
if both $Z_{1}$ and $Y_{1}$ have the same kind of reactance. An
example of dispersion diagram with a DBE (a fourth order EPD) at $k=0$
and a regular band edge (a second order EPD) at $k\neq0$, is shown
in Fig. 1(d), using the CTL parameters provided in the next section.
The DBE occurring at $k=0$, which is the main focus of this paper,
has a dispersion characterized by the relation \cite{F3,2017theory}

\begin{equation}
(\omega-\omega_{e})=hk^{4}
\end{equation}
in the vicinity of $k=0$ , where $h$ is a geometry-dependent fitting
parameter that controls the flatness of the dispersion.

Using the matrix notation as in \cite{paul2008analysis} for the circuit
\textit{equivalent} model in Fig. \ref{fig:CCT_Model}, the differential
wave equations (telegrapher's equations) describing propagation in
the two CTLs are

\begin{equation}
\begin{array}{c}
\begin{array}{c}
\dfrac{d\mathbf{V}(z)}{dz}=-\mathbf{\underline{\underline{Z}}}(\omega)\mathbf{I}(z),\\
\mathbf{}
\end{array}\\
\dfrac{d\mathbf{I}(z)}{dz}=-\mathbf{\underline{\underline{Y}}}(\omega)\mathbf{V}(z).
\end{array}\label{eq:telegraphic}
\end{equation}
Here $\mathbf{\underline{\underline{Z}}}$ and $\mathbf{\underline{\underline{Y}}}$
are the per-unit-length series-impedance and shunt-admittance matrices,
respectively, describing the per-unit-length distributed parameters
of the coupled transmission lines (CTLs) \cite{paul2008analysis}.
They are $2\times2$ symmetric matrices given by

\begin{equation}
\begin{array}{c}
\mathbf{\underline{\underline{Z}}}(\omega)=\left(\begin{array}{cc}
Z_{1}(\omega) & 0\\
0 & Z_{2}(\omega)
\end{array}\right),\\
\\
\mathbf{\underline{\underline{Y}}}=\left(\begin{array}{cc}
Y_{1}(\omega)+Y_{c}(\omega) & -Y_{c}(\omega)\\
-Y_{c}(\omega) & Y_{2}(\omega)+Y_{c}(\omega)
\end{array}\right),
\end{array}\label{eq:Z_Y}
\end{equation}
where the coupling between the two TLs is due to $Y_{c}(\omega)$.
For the sake of convenience, a four-dimensional state vector that
includes voltages and currents at a coordinate $z$ in the CTLs is
defined as 

\begin{equation}
\mathbf{\mathbf{\Psi}}(z)=[\begin{array}{cccc}
V_{1}(z)\ , & V_{2}(z)\ , & I_{1}(z)\ , & I_{2}(z)\end{array}]^{\mathrm{T}}.
\end{equation}
Therefore, the two telegrapher equations (\ref{eq:telegraphic}) representing
wave propagation are cast in terms of a multidimensional first order
differential equation \cite{othman2016theory}, \cite{yakovlev1998analysis}

\begin{equation}
\begin{array}{c}
\dfrac{d\mathbf{\mathbf{\Psi}}(z)}{dz}=-j\mathbf{\underline{M}}\mathbf{\mathbf{\mathbf{\mathrm{(\omega)}\Psi}}}(z),\end{array}\label{eq:telegraphic-1}
\end{equation}
where $\mathbf{\underline{M}}(\omega)$ is a $4\times4$ system matrix
given by

\begin{equation}
\mathbf{\underline{M}\mathrm{(\omega)}}=\left(\begin{array}{cc}
\mathbf{\underline{\underline{0}}} & -j\mathbf{\underline{\underline{Z}}}(\omega)\\
-j\mathbf{\underline{\underline{Y}}}(\omega) & \mathbf{\underline{\underline{0}}}
\end{array}\right),
\end{equation}
and $\mathbf{\underline{\underline{0}}}$ is the 2\texttimes 2 null
matrix.

\begin{figure}
\centering \subfigure[]{\includegraphics[width=0.24\textwidth]{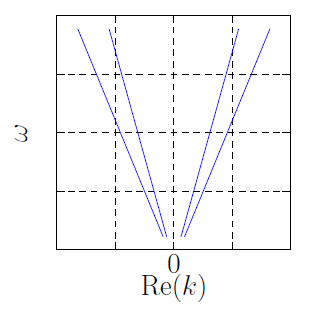}\label{fig:CTL_Uni-1}}\subfigure[]{\includegraphics[width=0.24\textwidth]{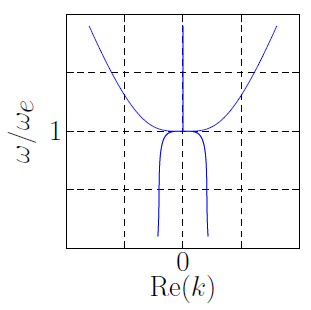}\label{fig:TL_Model-1}}

\centering \subfigure[]{\includegraphics[width=0.24\textwidth]{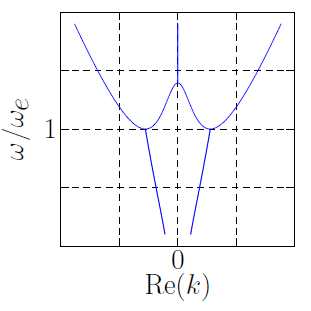}\label{fig:CCT_Model-2}}\subfigure[]{\includegraphics[width=0.24\textwidth]{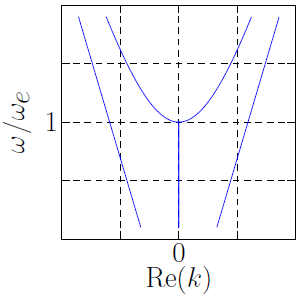}
\label{fig:CCT_Model-1-1}}
\raggedright{}\caption{Dispersion diagrams describing different EPDs at angular frequency
$\omega_{e}$: (a) CTLs where none of the EPD conditions are satisfied
at any non-zero frequency; (b) CTLs exhibiting the $4^{th}$ order
EPD (i.e., the DBE) at the angular frequency $\omega_{e}$ where $\mathrm{Tr}(\mathbf{\underline{\underline{Z}}\mathrm{(\omega_{e})}}\mathbf{\underline{\underline{Y}}\mathrm{\mathrm{(\omega_{e})}}})=0$
and $\mathrm{det}(\mathbf{\underline{\underline{Z}}(\omega_{e})\ }\mathbf{\underline{\underline{Y}}(\omega_{e})})=0$.
For uniform CTLs made of 2 TLs this condition necessarily occurs at
$k=0$; (c) CTLs exhibiting two exceptional points of $2^{nd}$ order
degeneracy, where $\mathrm{Tr}(\mathbf{\underline{\underline{Z}}\mathrm{(\omega_{e})}}\mathbf{\underline{\underline{Y}}\mathrm{\mathrm{(\omega_{e})}}})^{2}=4\mathrm{det}(\mathbf{\underline{\underline{Z}}(\omega_{e})\ }\mathbf{\underline{\underline{Y}}(\omega_{e})})$.
For uniform CTLs made of 2 TLs this can occur at any $k$; (d) CTLs
exhibiting a single $2^{nd}$ order EPD where $\mathrm{det}(\mathbf{\underline{\underline{Z}}(\omega_{e})\ }\mathbf{\underline{\underline{Y}}(\omega_{e})})=0$.
This condition occurs at $k=0$. In these plots we show only the real
part of the four modal wavenumbers.}
\label{fig:EPD_Curves}
\end{figure}
When the matrix $\underline{\mathbf{M}}(\omega)$ is diagonalizable
all the four eigenmodes supported in the CTL have state vectors $\mathbf{\Psi_{\mathit{n}}}(z)\propto e^{-jk_{n}z}$,
with $n=1,2,3,4$, see proof in Appendix \ref{sec:General-solution-of};
however, when the matrix $\underline{\mathbf{M}}(\omega)$ is not
diagonalizable (this is corresponding to the case exhibiting an EPD),
some modes preserve the proportionality $\mathbf{\Psi_{\mathit{n}}}(z)\propto e^{-jk_{n}z}$,
while the rest have algebraic growth with $z$ as $\mathbf{\Psi}_{n}\propto\mathbf{P}(z)e^{-jk_{n}z}$,
where $\mathbf{P}(z)$ is a vector polynomial function of maximum
order $3$ for systems made of two CTLs as considered in this paper,
see proof in Appendix \ref{sec:General-solution-of}. Therefore, when
$\underline{\mathbf{M}}(\omega)$ is diagonalizable the eigenmodes
supported by the uniform CTL described by (\ref{eq:telegraphic-1})
are fully represented by using $\mathbf{\Psi}(z)\propto e^{-jkz}$
in (\ref{eq:telegraphic-1}) to obtain $-jk\begin{array}{c}
\mathbf{\mathbf{\Psi}}(z)=-j\mathbf{\underline{M}}\mathbf{\mathbf{\mathbf{\mathrm{(\omega)}\Psi}}}(z)\end{array}$ \cite{othman2016theory}, yet simplified to an eigenvalue problem
as

\begin{equation}
\mathbf{\underline{M}}\mathbf{\mathbf{\mathbf{\Psi}}}(z)=k\mathbf{\mathbf{\mathbf{\Psi}}}(z).\label{eq:Eig_Prop}
\end{equation}

The four eigenvalues $k_{1},$ $k_{2}$, $k_{3}$ and $k_{4}$ and
their corresponding eigenvectors (at $z=0$) $\mathbf{\mathbf{\Psi}}_{1}$,
$\mathbf{\Psi}_{2}$, $\mathbf{\Psi}_{3}$ and $\mathbf{\Psi}_{4}$
of the above eigenvalue problem are determined as in Appendix \ref{sec:Solution-of-Eigenvalue}
and they are written in their simplest form as \cite{yakovlev1998analysis}
\begin{equation}
\begin{array}{c}
k_{1}=-k_{3}=\dfrac{1}{\sqrt{2}}\sqrt{-T-\sqrt{T^{2}-4D}},\\
k_{2}=-k_{4}=\dfrac{1}{\sqrt{2}}\sqrt{-T+\sqrt{T^{2}-4D}},
\end{array}\label{eq:dis}
\end{equation}
where $T=\mathrm{Tr}(\mathbf{\underline{\underline{Z}}}\ \mathbf{\underline{\underline{Y}}})$
is the trace and $D=\mathrm{det}(\mathbf{\underline{\underline{Z}}\ }\mathbf{\underline{\underline{Y}}})$.
The system vector is concisely and conveniently represented as

\begin{equation}
\mathbf{\Psi}_{n}=\psi_{0}\left(\begin{array}{c}
\begin{array}{c}
Z_{1}\left(k_{n}^{2}+Z_{2}(Y_{2}+Y_{c})\right)\\
Z_{1}Z_{2}Y_{c}
\end{array}\\
jk_{n}\left(k_{n}^{2}+Z_{2}(Y_{2}+Y_{c})\right)\\
jZ_{1}k_{n}Y_{c}
\end{array}\right),\label{eq:Eig_Ve}
\end{equation}
where $\psi_{0}$ is arbitrary constant and it has a unit of $\mathrm{Am^{3}}$.

\begin{figure}
\centering \subfigure[]{\includegraphics[width=0.23\textwidth]{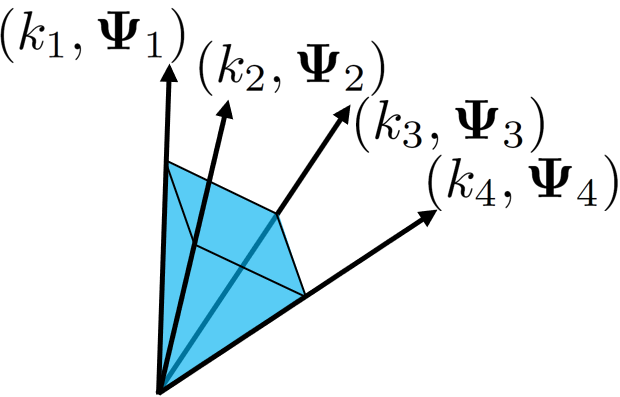}}
\subfigure[]{\includegraphics[width=0.23\textwidth]{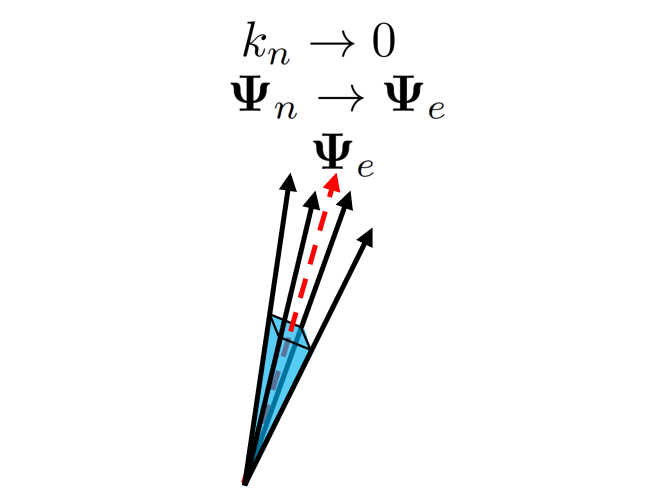}\label{fig:Four_EPD}}

\centering \subfigure[]{\includegraphics[width=0.23\textwidth]{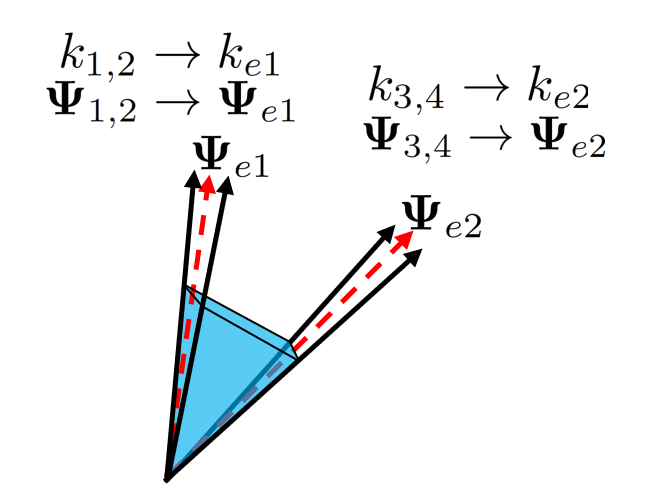}\label{fig:Four_EPD-1}}
\subfigure[]{\includegraphics[width=0.23\textwidth]{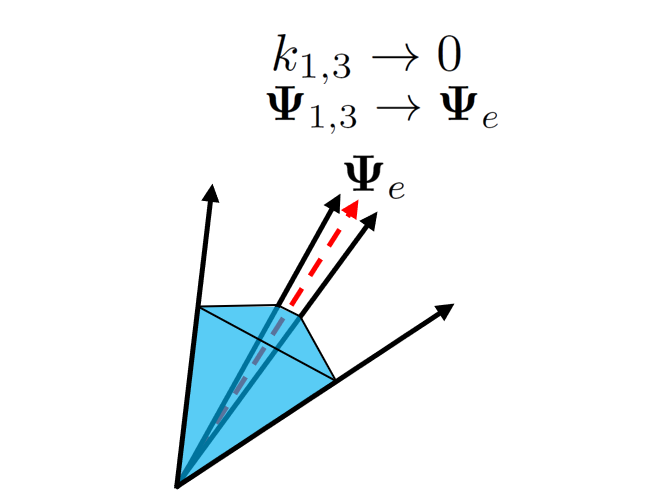}
}
\raggedright{}\caption{Schematic representation of the four eigenvectors of the four eigenmodes
supported by a CTL as they approach different EPDs conditions: (a)
no EPD, i.e., the four eigenvectors are four independent vectors in
a four dimensional state space; (b) $4^{th}$ order EPD, i.e., the
four eigenvectors tend to coalesce into a single eigenstate. When
the structure is lossless and gainless such $4^{th}$ order EPD is
called DBE; (c) two points of $2^{nd}$ order degeneracy, i.e, pairs
of eigenvectors coalesce to two independent eigenstates at the so
called RBE; and (d) a single $2^{nd}$ order EPD, i.e., only two eigenvectors
coalesce at the so called RBE while the other two remain independent.
The degree (i.e., the order) of degeneracy of a multimode EPD condition
is given by the number of coalescing eigenvectors. }
\label{fig:EPDS}
\end{figure}
The solutions (\ref{eq:dis}) and (\ref{eq:Eig_Ve}) represent the
four wavenumbers of the eigenmodes that propagate or attenuate along
both the positive and negative \textit{z}-directions (four modes),
viz., $k_{3}=-k_{1}$ and $k_{4}=-k_{2}$. 

In general, an EPD of order $m$ occurs when $m$ eigenomodes have
the same eigenvalue and eigenvector. For a system of two uniform CTLs
a $4^{th}$ order EPD (a full order EPD) occurs if all the $4$ eigenvalues
are equal \cite{yakovlev1998analysis}, which implies that eigenvectors
coalesce as well, as it is obvious from $(\ref{eq:Eig_Ve})$. Therefore
in such a uniform system the coalescence of four wavenumbers is a
sufficient condition for an EPD to occurr.

The system made of two CTLs considered in this paper exhibits three
types of EPDs: (i) two points of second order degeneracy ($k_{1}=k_{2}$
and $k_{3}=k_{4})$ when $\mathrm{\mathit{T}}{}^{2}=4D$. This can
occur at any wavenumber $k$; (ii) A second order EPD ($k_{1}=k_{3})$
or ($k_{2}=k_{4})$ when $D=0$. This occurs only at $k=0$; and (iii)
a fourth order EPD $(k_{1}=k_{2}=k_{3}=k_{4})$ when both $T=0$ and
$D=0$. This occurs only at $k=0$. These three cases are illustrated
in the dispersion diagram in Fig. \ref{fig:EPD_Curves} and in the
schematic representation of the four eigenvectors in Fig. \ref{fig:EPDS}.
Indeed, in a reciprocal systems ($k_{1}=-k_{3}),$ the equality ($k_{1}=k_{3})$
in condition (ii) implies that ($k_{1}=k_{3}=0).$ Furthermore, still
based on reciprocity, the condition $(k_{1}=k_{2}=k_{3}=k_{4})$ in
(iii) implies that $(k_{1}=k_{2}=k_{3}=k_{4}=0)$. Hence, these two
conditions can be used also to design systems radiating at broadside
and working at an EPD. Condition (ii) is usually refereed to as a
cutoff condition (at $k=0$) and indeed it occurs also in regular
single mode waveguides. Condition (i) is interesting, because it sets
a cutoff condition at any desired wavenumber \textcolor{black}{$k\neq0$.
It is important to point out that a $3^{rd}$ order EPD is not possible
to exist in two coupled transmission lines unless we break reciprocity
\cite{5340521} which is out of the scope of this paper; here we only
consider reciprocal coupled transmission lines. The scope of this
paper is mainly to show the fourth order degeneracy (namely the DBE)
described in condition (iii) and to show that condition (i) can be
also easily engineered.}

\section{Fourth Order DBE in Uniform Waveguides}

When modes are supported in uniform waveguides modeled by two uniform
and coupled TLs, a fourth order EPD (DBE) occurs when all four independent
eigenvectors coalesce and form one single eigenvector \cite{F3},
\cite{figotin2003oblique} as schematically shown in Fig. \ref{fig:Four_EPD}.
This occurs when the impedance and admittance matrices that describe
the per-unit-length parameters of the system satisfy both conditions:

\begin{equation}
\begin{array}{c}
\mathrm{\mathit{T}=Tr}(\mathbf{\underline{\underline{Z}}\ }\mathbf{\underline{\underline{Y}}})=0,\\
D=\mathrm{det}\mathbf{(\underline{\underline{Z}}}\ \mathbf{\underline{\underline{Y}}})=0.
\end{array}\label{eq:ZY_Cond}
\end{equation}

Indeed from (\ref{eq:ZY_Cond}) these two conditions imply that $k_{1}=k_{2}=k_{3}=k_{4}$
and consequently from (\ref{eq:Eig_Ve}) it implies that all four
eigenvectors are identical. Substituting (\ref{eq:Z_Y}) into (\ref{eq:ZY_Cond})
and after some simplification, necessary and sufficient conditions
to realize a fourth order EPD at radian frequency $\omega_{e}$ in
term of the per-unit-length CTL parameters are obtained in their simplest
form as

\begin{equation}
Z_{1}(\omega_{e})Y_{1}^{2}(\omega_{e})=-Z_{2}(\omega_{e})Y_{2}^{2}(\omega_{e}),\label{eq:C1}
\end{equation}

\begin{equation}
Y_{c}(\omega_{e})=\frac{-Y_{1}(\omega_{e})Y_{2}(\omega_{e})}{Y_{1}(\omega_{e})+Y_{2}(\omega_{e})}.\label{eq:C2}
\end{equation}

\begin{figure}
\centering \subfigure[]{\includegraphics[width=0.23\textwidth]{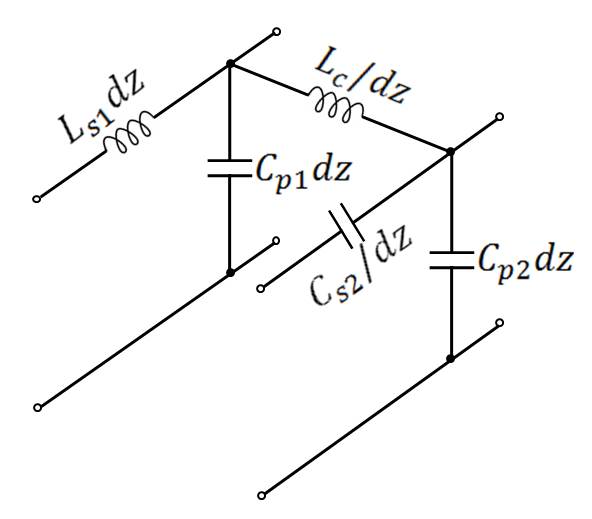}\label{fig:C1}}
\subfigure[]{\includegraphics[width=0.23\textwidth]{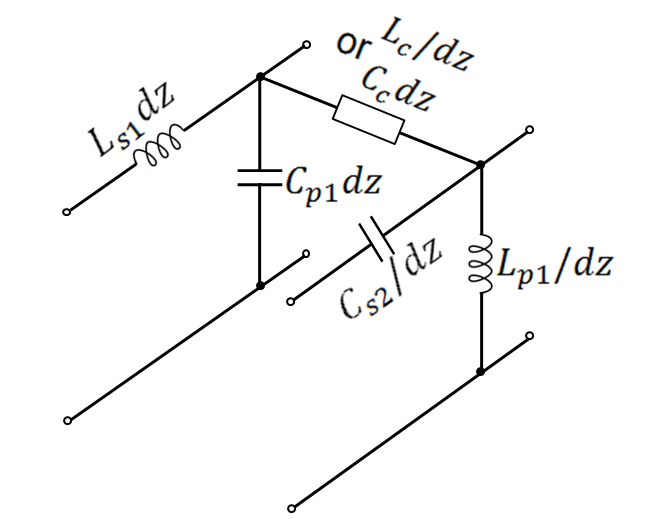}\label{fig:C2}}

\centering \subfigure[]{\includegraphics[width=0.23\textwidth]{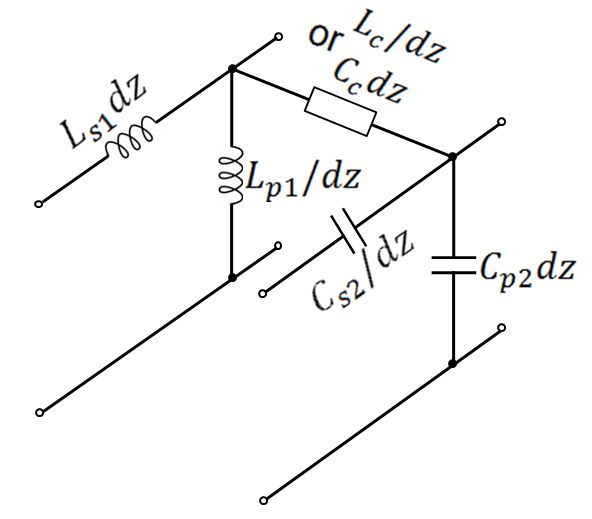}\label{fig:C3}}
\subfigure[]{\includegraphics[width=0.23\textwidth]{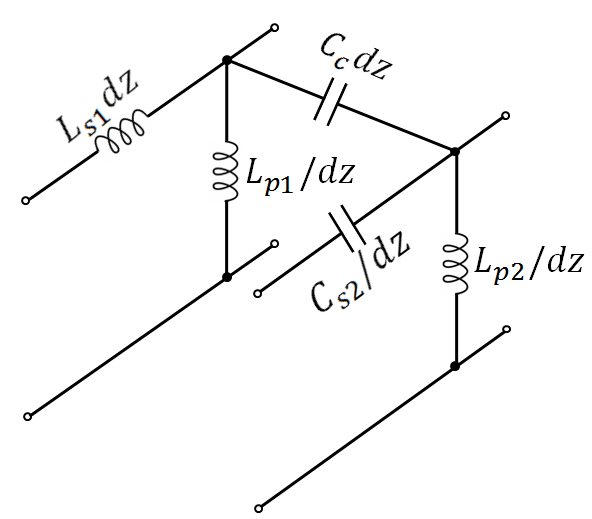}
\label{fig:C4}}
\raggedright{}\caption{Different configurations of uniform CTLs that may exhibit a fourth
order EPD, namely, the DBE when the CTLs are lossless. Here we show
the combinations of distributed reactances that provide multi-mode
degenerate conditions: Configuarion (a) shows that a fourth order
EPD is obtained by a proper inductive coupling between a ``forward''
propagating mode in $\mathrm{TL_{1}}$ and an evanescent mode in $\mathrm{TL_{2}}$.
Configuration (b) shows that a fourth order EPD is obtainable by a
proper coupling between a forward mode in $\mathrm{TL_{1}}$ and a
``backward'' mode in $\mathrm{TL_{2}}$. Note that here we denote
a mode to be ``forward'' when phase and group velocities have the
same signs, whereas a ``backward'' mode has phase and group velocities
with opposite signs. Configuration (c) shows that a four order EPD
is obtainable also when proper coupling is designed between evanescent
modes in $\mathrm{TL_{1}}$ and $\mathrm{TL_{2}}$. Finally, configuration
(d) shows that an EPD can be obtained also by a capacitive coupling
between an evanescent mode in $\mathrm{TL_{1}}$ and a backward propagating
mode in $\mathrm{TL_{2}}$.}
\label{fig:Inf_Conf}
\end{figure}

\begin{flushleft}
It is important to point out that the first condition in (\ref{eq:C1})
represents a constraint on the parameters of the uncoupled TLs to
have a DBE, whereas the second condition in (\ref{eq:C2}) represents
the constraint on the required coupling admittance to have a DBE.
Therefore just fixing the coupling parameter is not enough to have
a DBE since the two individual TLs (without considering coupling)
need to satisfy the constraint (\ref{eq:C1}). Both terms $Y_{1}^{2}(\omega_{e})$
and $Y_{2}^{2}(\omega_{e})$ in (\ref{eq:C1}) have a negative sign
(we do not consider losses so far in this ideal analysis) regardless
of the type of $Y_{1}$ and $Y_{2}$ susceptance. Consequently, from
(\ref{eq:C1}) and (\ref{eq:C2}) we deduce that two necessary conditions
to realize a fourth order EPD at radian frequency $\omega_{e}$ are 
\par\end{flushleft}

\begin{equation}
\begin{array}{c}
\mathrm{Im}\left(Z_{1}\right)\mathrm{Im}\left(Z_{2}\right)\Bigg|_{\omega=\omega_{e}}<0,\\
\\
\mathrm{Im}\left(Y_{c}^{-1}\right)\mathrm{Im}\left(Y_{1}^{-1}+Y_{2}^{-1}\right)\Bigg|_{\omega=\omega_{e}}<0.
\end{array}\label{eq:C1_F}
\end{equation}
This means that the a necessary condition to realize a DBE in uniform
CTL is that the two series per-unit-length impedances $Z_{1}$ and
$Z_{2}$ must be of different types, i.e., one should be capacitive
and the other inductive. Furthermore, $Y_{c}^{-1}$ and $Y_{1}^{-1}+Y_{2}^{-1}$
must also be of different types. Figure \ref{fig:Inf_Conf} shows
all possible configurations of the per-unit-length parameters of CTLs
that exhibit a fourth order DBE. 

From Fig. \ref{fig:Inf_Conf}, it is concluded that a fourth order
DBE occurs in two uniform CTLs when there is a coupling between: a
forward propagating mode and an evanescent mode (Fig. \ref{fig:C1}),
a forward and a backward propagating modes (Fig. \ref{fig:C2}), two
evanescent modes (Fig. \ref{fig:C3}), or a backward propagating mode
and an evanescent mode (Fig. \ref{fig:C4}). For a rectangular waveguide
structure, the configuration in Fig. \ref{fig:C1} represents a coupling
between a transverse electric (TE) or transverse magnetic (TM) propagating
mode and a TM evanescent mode (below cutoff), whereas the configuration
in Fig. \ref{fig:C3} represents coupling between TE and TM evanescent
modes, both below cutoff when considered without coupling \cite{Wave_Scenarios}.

\subsection{Example of uniform CTL with infinite length}

\begin{figure}
\begin{centering}
\includegraphics[width=0.96\columnwidth]{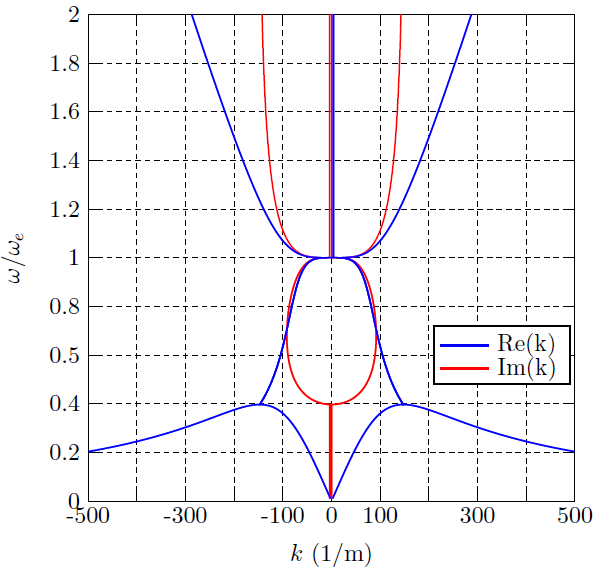}
\par\end{centering}
\caption{Dispersion diagram of modal complex wavenumbers $k$ versus normalized
frequency for two uniform CTLs with distributed circuit model as in
\ref{fig:C1}. The diagram shows a fourth order DBE $\omega=\omega_{e}$,
i.e., at $f=f_{e}=5\mathrm{GHz},$ where all modes have $k=0$. This
CTL structure also exhibits two RBEs (EPDs of second order) at $\omega=0.4\omega_{e}$,
i.e., at $f=2\mathrm{GHz}$, with a non-vanishing wavenumber of $k=\pm147.5\mathrm{m^{-1}}$.
The dispersion diagram showing \textit{only} the purely-real wavenumber
branches is reported in Fig. 1(d).}
\label{fig:case_1}
\end{figure}

Two CTLs with circuit configuration as in Fig. \ref{fig:C1} are designed
to exhibit a fourth order EPD at frequency $f_{e}=5\mathrm{\ GHz},$
i.e., to satisfy the DBE conditions in (\ref{eq:C1}) and (\ref{eq:C2}).
The CTLs parameters are $C_{p1}=C_{p2}=0.12\mathrm{\ nF/m}$, $L_{s1}=200\ \mathrm{nH/m}$,
$C_{s2}=5.07\mathrm{\ fFm}$ and $L_{c}=16.89\mathrm{\ pHm}$, where
the series and parallel per-unit-length components are designated
with subscripts $s$ and $p$, respectively. This is the case when
one TL (without considering the coupling between the two TLs) supports
two propagating modes (one in each direction) while the other TL supports
evanescent waves. However the two TLs are coupled via the inductive
subsceptance $Y_{c}=1/(j\omega L_{c}$) leading to the modal dispersion
diagram in Fig. \ref{fig:case_1}. There, both the real and imaginary
parts of the wavenumber are shown versus real radian frequency. A
fourth order DBE occurs at radian frequency $\omega_{e}=31.42\times10^{9}\ \mathrm{rad/s}$
at which $k_{1}=k_{2}=k_{3}=k_{4}=0$. Note that the dispersion diagram
also exhbits two second order EPDs which represent two regular band
edges (RBEs) at $\omega=0.4\omega_{e}$ (i.e., at $f\approx2\mathrm{GHz}$)
at two distinct non-vanishing wavenumbers $k=\pm147.5\mathrm{m^{-1}}$,
where their sufficient condition $\mathrm{T^{2}}=4D$ is satisfied
at this particular frequency. In the bandgap $0.4\omega_{e}<\omega<\omega_{e}$
the diagram has four wavenumbers with complex values and waves are
all evanescent. For $\omega>\omega_{e}$ two waves are propagating
and two are evanescent. The same dispersion diagram showing \textit{only}
the branches with purely-real wavenumbers is reported in Fig. 1(d).
Therefore the CTL technique used in this paper allows to put regular
band edges at properly designed wavenumbers.

\subsection{Uniform waveguide with finite length}

\begin{figure}
\begin{centering}
\centering \subfigure[]{\includegraphics[width=1\columnwidth]{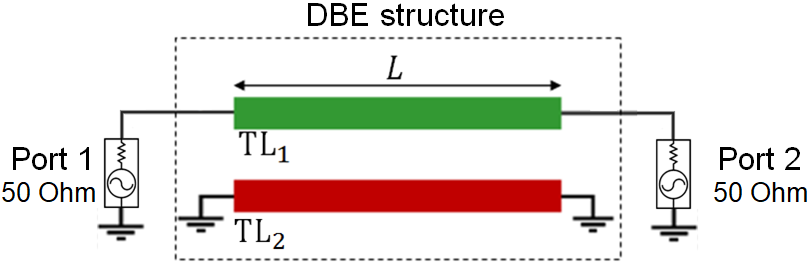}\label{fig:Q_1}} 
\par\end{centering}
\begin{centering}
\centering \subfigure[]{\includegraphics[width=0.94\columnwidth]{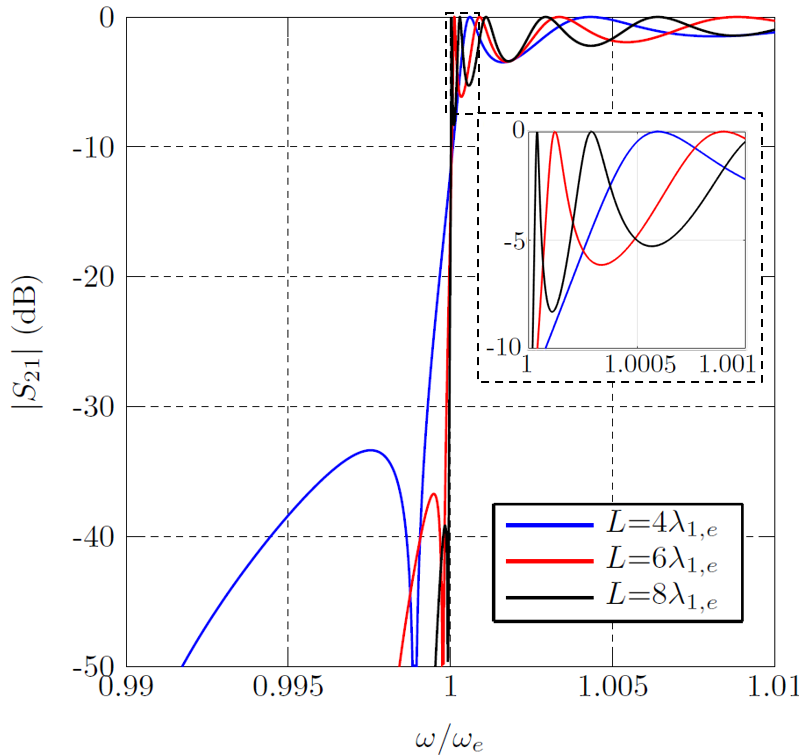}\label{fig:Q_2}} 
\par\end{centering}
\caption{\textcolor{black}{Magnitude of the transmission scattering parameter
$S_{21}$ for the waveguide consisting of two uniform microstrip CTLs
with finite length }\textit{\textcolor{black}{L,}}\textcolor{black}{{}
with distributed circuit model as in \ref{fig:C1}. The CTLs have
a fourth order EPD (namely, a DBE) at the so called DBE frequency
$f=f_{e}=5\mathrm{GHz}$ . (a) Finite length CTL circuit setup. (b)
Scattering parameter $S_{21}$ for different lengths }\textit{\textcolor{black}{L}}\textcolor{black}{{}
revealing that this finite length CTL structure is a cavity despite
the characteristic impedance of $\mathrm{TL_{1}}$ is equal to the
termination load. A clear transmission peak, called DBE resonance,
is observed near the DBE frequency, and it gets narrower for increasing
lengths. $\lambda_{1,e}$ is the wavelength of the propagating waves
in $\mathrm{TL}_{1}$, when it is uncoupled to $\mathrm{TL}_{2}$,
calculated at the EPD frequency $\lambda_{1,e}=2\pi/k_{1,e}=40.8$
mm, where $k_{1,e}=\omega_{e}\sqrt{L_{s1}C_{p1}}$.}}
\label{fig:case_1-1}
\end{figure}

So far we ha\textcolor{black}{ve discussed modal propagation in infinitely
long structures. We now consider two uniform CTLs with finite length
$L$, Fig. \ref{fig:Q_1}, operating in very close proximity of the
DBE, and investigate the transmission properties in terms of scattering
parameter $|S_{21}|$. Since this finite length CTL structure forms
a resonator, we also investigate its quality factor. The CTL per-unit-length
parameters are the same as those used in the previous subsection that
led to Fig. \ref{fig:case_1}. The top TL is connected with two ports,
left and right, whereas the bottom one is terminated on short circuits
at both ends, as depicted in Fig. \ref{fig:Q_1}. Figure \ref{fig:Q_2}
shows the transmission coefficient magnitude $|S_{21}|$ versus frequency,
for different lengths $L$. The length is here givenin terms of wavelengths
of the propagating wave in $\mathrm{TL}_{1}$, whenuncoupled to $\mathrm{TL}_{2}$,
calculated at the EPD frequency $\lambda_{1,e}=2\pi/k_{1,e}$= 40.8
mm, where $k_{1,e}=\omega_{e}\sqrt{L_{s1}C_{p1}}$. The pass band
property is in agreement with that shown in Fig. \ref{fig:case_1},
i.e., there is propagation for $f>f_{e}=5\mathrm{GHz}$. It is shown
that the CTL exhibits a resonance (called DBE resonance) at a frequency
almost coincident with the DBE one, regardless of the CTL length,
at least for the two longer cases. The frequency of the other resonances
at lower frequencies are strongly affected by the length of the structure.
This resonator based on a multi-mode degeneracy exhibits a very interesting
physical behavior of its quality factor. The loaded quality fa}ctor
of the finite length and lossless CTL is plotted versus length $L$
in Fig. \ref{fig:QF}, and it is concluded that such quality factor
(blue line) follows the asymptotic trend proportional to $L^{5}$
as $L$ increases, which is the same conclusion that was made in \cite{sloan2017theory},
\cite{F2}, \cite{nada2018giant} and \cite{noh2010giant}, though
in these references the DBE was obtained in periodic structures and
at the edge of the Brillouin zone, whereas in this paper we show for
the first time a DBE at \textit{k}=0. Here the quality factor has
been evaluated as $Q=\omega_{res}\tau_{g}/2$, where $\omega_{res}$
is the resonance frequency associated with maximum transmission, i.e.,
where $|S_{21}|$ is maximum, and the group delay $\tau_{g}$ is calculated
as the derivative of the phase of $S_{21}$, with respect to the angular
frequency $\omega$, i.e., $\tau_{g}=\text{\ensuremath{\partial}}\left(\text{\ensuremath{\angle}}S_{21}\right)/\text{\ensuremath{\partial}}\omega$
\cite{Quality_Ref}. Note that high \textit{Q} values are obtained
while the $\mathrm{TL_{1}}$ characteristic impedance (without considering
the coupling) is 50 Ohms and the termination load is also 50 Ohms;
therefore TLs forming a cavity using the four mode degenerate condition
(the DBE) do not need high reflection coefficients at the end of each
TL. The strong reflection of the degenerate modes at the end of the
CTL occurs because the characteristic impedance of a CTL made of two
TLs is actually represented by a 2$\times$2 impedance matrix, and
therefore it is generally mismatched when two independent loads are
used as termination as in Fig. \ref{fig:Q_1}. Furthermore, exactly
at the DBE frequency the group velocities of the four coalescing modes
vanish and therefore the characteristic 2$\times$2 matrix impedance
shall describe absence of power flow (the characteristic impedance
of a \textit{single} TL at cutoff would be either zero or infinity).
However one should note that the DBE \textit{resonance} is slightly
shifted from the DBE frequency and therefore power transfer to the
load is actually occurring. It is important to point out that there
are various resonance frequencies in the cavity, however, in this
paper we are focusing on the nearest one to the DBE frequency which
we call it the first resonance frequency. Because of the DBE-like
dispersion relation in (2), for long cavities the first DBE resonance
frequency is approximated by the asymptotic formula

\begin{equation}
f_{res,1}=f_{e}+\alpha/L^{4},
\end{equation}
where $\alpha$ is a constant. This implies that the longer the CTL
cavity, the closer the DBE resonance is to the DBE frequency, and
hence the less power leakage occurs outside the resonator.

A further investigation is now conducted by studying the effect of
series and parallel distributed losses in the CTL on the quality factor.
Therefore we assume that each TL has either a per-unit-length series
resistance $R_{s}$ or a per-unit-length shunt conductance $G_{p}$.
Accordingly, Fig. \ref{fig:QVD_R} plot the quality factor of the
CTL versus length $L$ for different values of the series quality
factor $Q_{s}$, where $Q_{s}=\omega_{e}L_{s1}/R_{s}=1/(\omega_{e}C_{s2}R_{s})$
is the quality factor (assumed the same) of the two series elements,
which are an inductive distributed reactance in $\mathrm{TL_{1}}$
and a capacitive distributed reactance in $\mathrm{TL_{2}}$, and
hence they satisfy $\omega_{e}L_{s1}=1/(\omega_{e}C_{s2})$. In Fig.
\ref{fig:QVD_G} instead we show the quality factor by considering
losses in the two shunt (parallel) capacitive susceptances such that
$Q_{p}=\omega C_{p1}/G_{p1}=\omega C_{p2}/G_{p2}$. Note that the
same parallel capacitor and same loss is used in each of the two TLs.
The two plots show a very important fact about \textit{uniform} CTLs
exhibiting a fourth order DBE: the quality factor of the CTLs is robust
to the series losses, i.e., the series distributed resistance does
not affect the total quality factor trend shown in Fig. \ref{fig:QVD_R}.
This occurs because the wavenumbers of the four modes at DBE are such
that $k_{1}=k_{2}=k_{3}=k_{4}=0$, which means the voltage along the
finite length CTL is basically constant resulting in an almost vanishing
current through the series elements $Z_{1}$ and $Z_{2}$ . However,
when losses are in the shunt (parallel) elements the quality factor
of the structure tends to saturate to the quality factor of the used
distributed parallel capacitors as shown in Fig. \ref{fig:QVD_G}.
To obtain such plots, for each CTL length we have determined the resonant
frequency and evaluated the required parameters at that frequency.

It is important to point out that the resonance mentioned in the previous
study is not a conventional resonance due to two mode reflection,
however, it is due to four modes which make it with very unique properties
like quality factor and resonance frequency scaling with cavity length.
Such properties can be used to make oscillator with a unique mode
selection scheme that leads to a stable single-frequency oscillation,
even in the presence of load variation \cite{abdelshafy2020distributed,oshmarin2016oscillator}.
Moreover, the proposed DBE in this paper exists at $k=0$ which make
good candidate for application like leaky wave antennas, and active
leaky wave antennas that act as radiating oscillators.

\begin{center}
\begin{figure}
\begin{centering}
\centering \subfigure[]{\includegraphics[width=0.8\columnwidth]{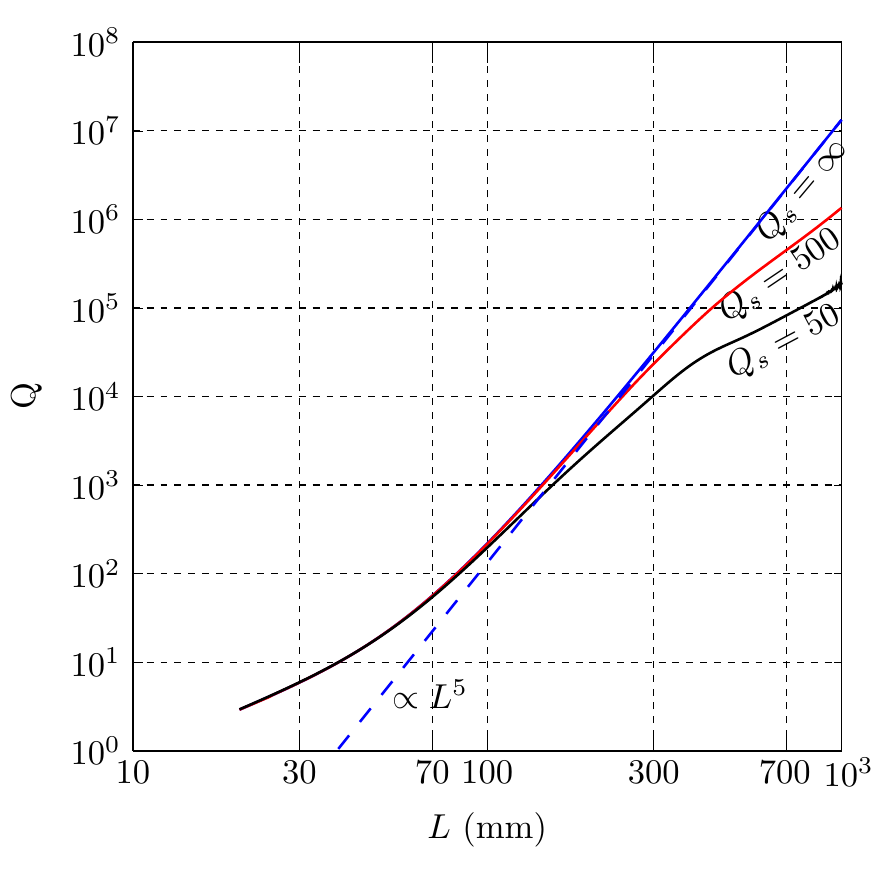}\label{fig:QVD_R}} 
\par\end{centering}
\begin{centering}
\centering \subfigure[]{\includegraphics[width=0.8\columnwidth]{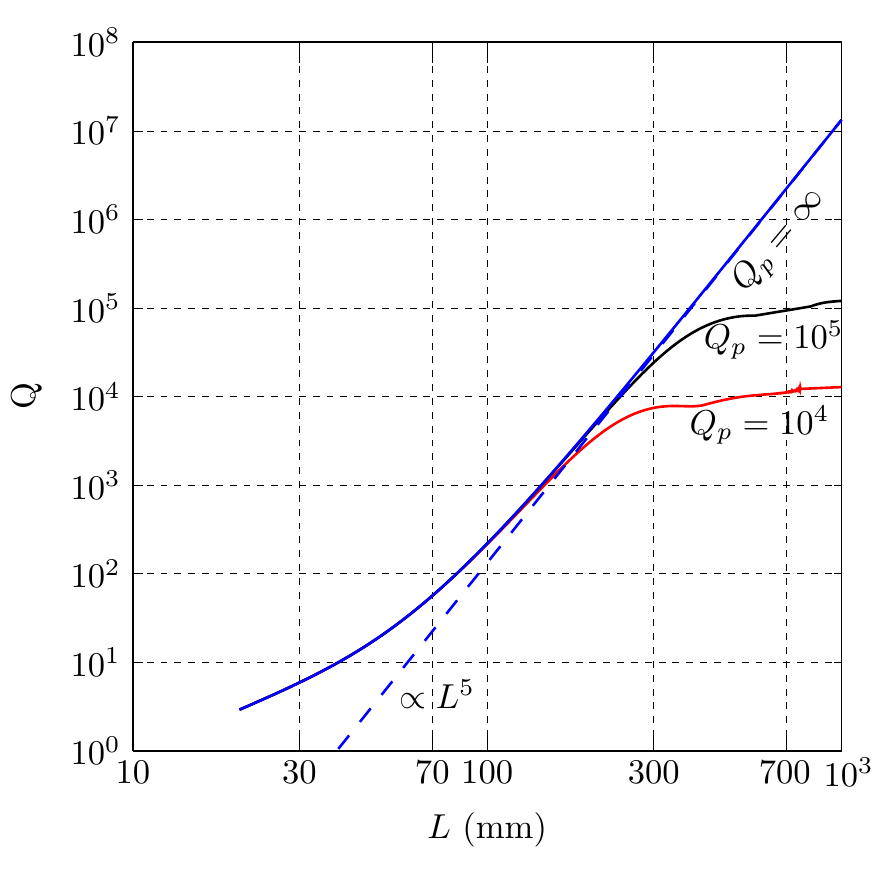}\label{fig:QVD_G}} 
\par\end{centering}
\centering{}\caption{Trend of the quality factor of a CTL cavity as in Figure \ref{fig:Q_1}
operating at the DBE resonance, in close proxinity of the DBE frequency,
showing the $L^{5}$ scaling with cavity length $L$. When the CTL
cavity has distributed losses, the quality factor trend is perturbed.
Distributed series resistance and parallel conductance are assumed
to be symmetrical, i.e., identical in each TL: (a) series losses only,
and (b) parallel losses only. The legend $Q=\infty$ refers to the
limit represented by a lossless CTL cavity and the blue dashed line
is a fitting trend showing the $L^{5}$ growth with cavity length.
These plots show that the Q factor of the CTL cavity is less sensitive
to series losses. }
\label{fig:QF}
\end{figure}
\par\end{center}

\section{Microstrip Implementation with Subwavelength Series Capacitors}

\begin{figure}
\centering{}\includegraphics[width=0.85\columnwidth]{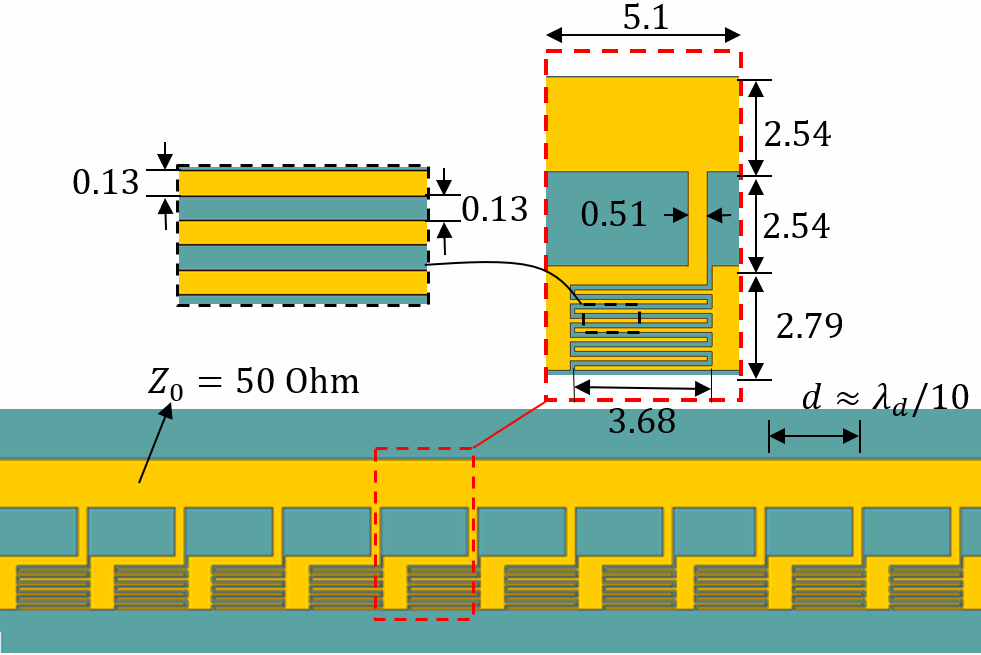}\caption{Microstrip implementation of two uniform CTLs over a grounded dielectric
substrate, with circuit model as in Fig. \ref{fig:C1}, i.e., with
a distributed series capacitor (bottom line) that is here implemented
by resorting to a periodic distribution of series inter-digital capacitors,
with sub-wavelength period $d$. The bottom part of the figure shows
the finite length CTLs, whereas the top part shows the unit cell with
period $d=$5.1\ mm. Dimensions are all in mm. This microstrip CTL
implementation develops a fourth order EPD at $f=f_{e}=5\mathrm{GHz}$.}
\label{fig:Micro_imp}
\end{figure}

A microstrip implementation of the uniform CTL in Fig. \ref{fig:C1}
is now considered where the series continuously distributed capacitance
is approximated by a periodic capacitive loading with subwavelength
period $d=\lambda_{d}/10$ ($\lambda_{d}$ is wavelength in the substrate,
that approximates the guided wavelength in a single microstrip line).
The very subwavelegth period makes the structure approximately uniform
following the metamaterial TLs concepts. Indeed we design the CTL
such that the homogenized effective CTL parameters approximately equal
those in the uniform case considered in the previous subsections.
The grounded dielectric substrate has a relative dielectric constant
of 2.2, loss tangent 0.001, and height of 0.75 mm. Metal layers have
conductivity of $4.5\times10^{7}\ $S/m and thickness of $35\ \mathrm{\mu m}$.
The series capacitance in each unit cell is implemented using an inter-digital
capacitor and the coupling inductance in Fig. \ref{fig:C1} is implemented
using a folded short and thin microstrip between the two TLs as shown
in Fig. \ref{fig:Micro_imp}. The two TL widths (i.e., when assumed
uncoupled, and before introducing the series capacitors) are designed
to have a characteristic impedance of 50 Ohms at $f=$5\ GHz. All
the dimensions (in mm) are reported in Fig. \ref{fig:Micro_imp}.
The inter-digital capacitance is approximately $C_{d}=$1\ pF, and
since the period is $d=5.1$\ mm, then the effective distributed
series capacitance is the same as the required one to get DBE, i.e.,
$C_{s1}=C_{d}d\approx5.1\ $fFm.

\begin{figure}
\begin{centering}
\centering \subfigure[]{\includegraphics[width=0.8\columnwidth]{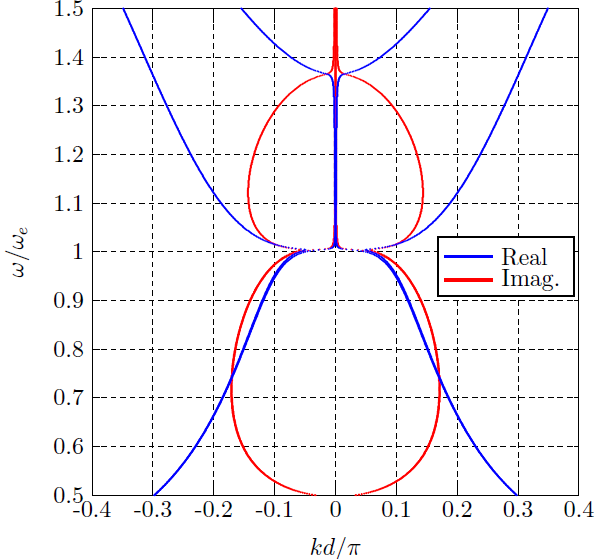}\label{fig:Dis_Mic}} 
\par\end{centering}
\begin{centering}
\centering \subfigure[]{\includegraphics[width=0.8\columnwidth]{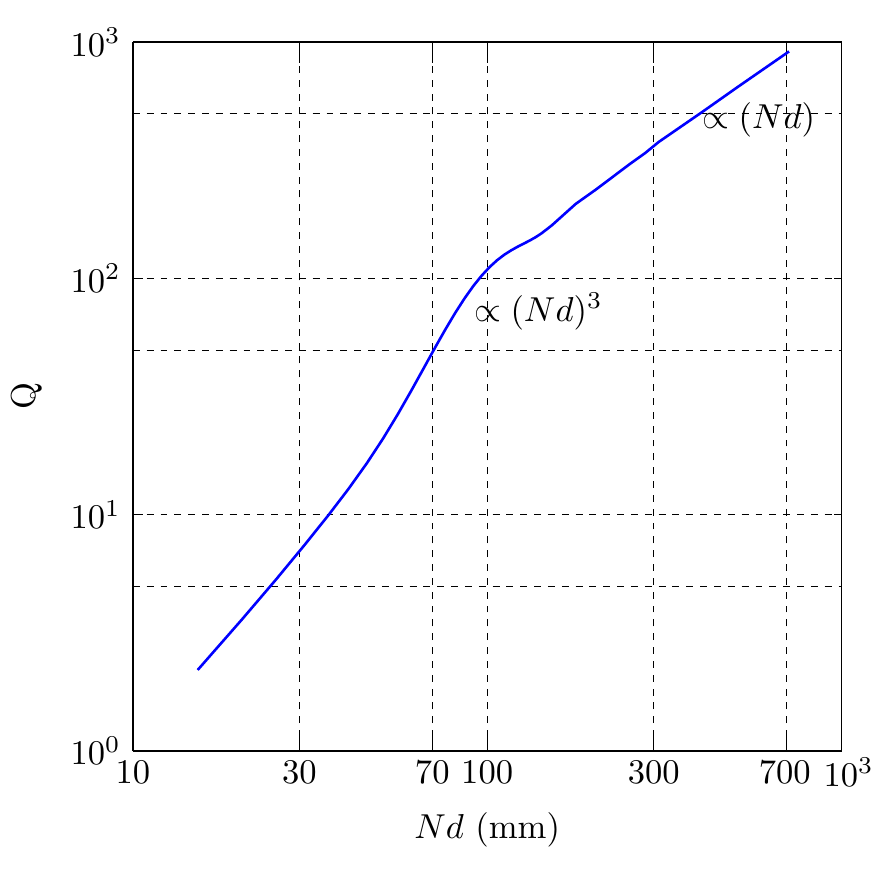}\label{fig:Dis_Mic-1}}
\par\end{centering}
\caption{Results relative to the microstrip implementation of the uniform CTLs
using a periodic distribution of interdigital series capacitors in
the bottom TL, with subwavelength period (Fig. \ref{fig:Micro_imp}).
(a) Dispersion diagram obtained via full-wave simulation showing the
complex modal wavenumbers versus frequency. The full-wave simulation
reveals the existence of a DBE (a fourth order degeneracy) at \textit{\textcolor{black}{k}}
= 0. The simulation accounts for radiation, dielectric and copper
losses. (b) Quality factor of the periodic CTLs versus resonant ``cavity''
length, showing its scaling with the number of unit cells \textit{\textcolor{black}{N}}.}
\end{figure}

Figure \ref{fig:Dis_Mic} shows the modal dispersion obtained using
full wave simulations based on the method of moments implemented in
Keysight Technologies Advanced Design System (ADS). The used method
of moments is based on the three-dimensional Green's function with
all the dynamic terms, hence including radiation losses. The dispersion
relation was calculated by determining the S-parameters of a single
unit-cell, then converting them to a 4$\times$4 unit-cell transfer
matrix $\mathbf{T}_{U}$ that relates voltages and currents at the
beginning and end of the unit cell as in \cite{abdelshafy2018exceptional},
and then using the Floquet theorem determining the eigenvalue problem
that provides the four modal wavenumbers (see also Appendix A). Figure
\ref{fig:Dis_Mic} shows the existence of a DBE in the dispersion
diagram, and in proximity of $\omega_{e}$ it is in good agreement
with the diagram of the uniform ideal CTL in Fig. \ref{fig:case_1}.

We then observe the quality factor of a resonator made by a finite-length
dual microstrip, shown in Fig. \ref{fig:Micro_imp}. The loading and
excitation for calculating the quality factor are as shown in Fig.
\ref{fig:Q_1} , and the operating frequency is at the DBE resonance
(the peak of the transfer function closest to the DBE frequency).
The quality factor is estimated by the same formula considered in
the previous section, i.e., by $Q=\omega_{res}\tau_{g}/2$, where
the resonance frequency (the one closest to the DBE frequency) depends
on the cavity length. The quality factor versus ``cavity'' length
$L=Nd$ , using $N$ unit cells of the microstrip implementation in
Fig. \ref{fig:Micro_imp}, is plotted in Fig. \ref{fig:Dis_Mic-1}.
From this figure we note that the quality factor tends to saturate
before exhibiting the asymptotic $L^{5}$ trend because of radiation,
conduction and dielectric losses. Indeed the ideal $Q\propto L^{5}$
trend depicted in Fig. 7 (blue line) occurs only in the ideal case
where losses are negligible, whereas in this case both series and
shunt losses are present because of copper and dielectric losses.
Note that here the $\mathrm{TL_{1}}$ characteristic impedance is
50 Ohms and that the load is also 50 Ohms, therefore a cavity using
the four mode degenerate condition (the DBE) does not need high reflection
coefficients at the end of each TL that can be normally terminated
at any load.


\section{\textcolor{black}{Experimental Verification Using a CTL with Discrete
Series capacitor}}

\begin{figure}
\centering{}\textcolor{black}{\includegraphics[width=0.72\columnwidth]{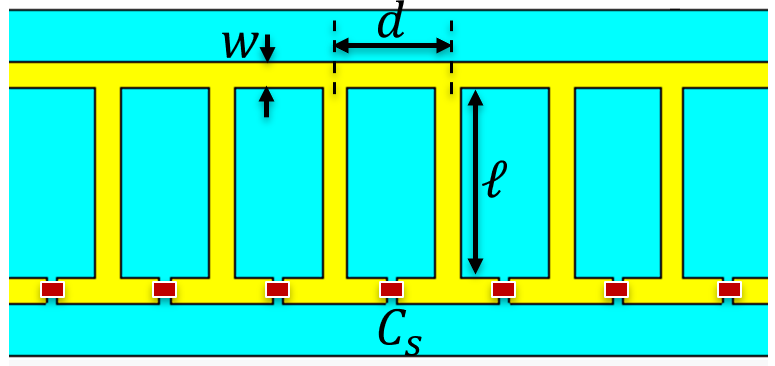}\caption{\textcolor{black}{Microstrip implementation of a waveguide made of
two coupled uniform TLs over a grounded dielectric substrate in Fig.
\ref{fig:C1} that exhibits DBE. The top TL (when uncoupled to the
bottom one) supports propagation. The bottom TL (when uncoupled to
the top one) supports evanescent modes because it is loaded with distributed
series capacitors mimicking a uniform series capacitive per-unit-length
distribution. The capacitors in this structure are discrete components
with value $3.1$\ pF. The inductive coupling between the two TLs
is implemented using stubs connected between the top and the bottom
TLs. The period is small compared to the guided wavelength.}}
\label{fig:Micro_imp-1}}
\end{figure}

\textcolor{black}{In this section we show an experimental verification
of the existence of the DBE when and evanescent modes are coupled
in the CTL. Figure \ref{fig:Micro_imp-1} shows the microstrip implementation
of the uniform CTL in Fig. \ref{fig:C1}. The unit-cell is fabricated
on a grounded dielectric substrate (Rogers substrate RT/duroid 5880)
with a relative dielectric constant of 2.2, loss tangent of 0.001,
and height of 0.79\ mm. We use here discrete component capacitors
to periodically load one TL to support evanescent modes. We use surface
mount ceramic capacitors (manufactured by Murata Electronics, part
number GJM1555C1H3R1BB01D) with capacitance of $3.1$\ pF and quality
factor of $Q>50$ for $f<3$\ GHz. All the TLs have width of $w=2.4$\ mm
to have a characteristic impedance of $50$\ Ohm. The structure has
period of $d=10.5$\ mm ($d\sim\lambda_{d}/10$) and stubs length
$\ell=19$\ mm. As discussed in the previous section, the CTL can
be seen as uniform, due to the subwavelength period.}

\begin{figure}
\begin{centering}
\centering \subfigure[]{\textcolor{black}{\includegraphics[width=0.225\textwidth]{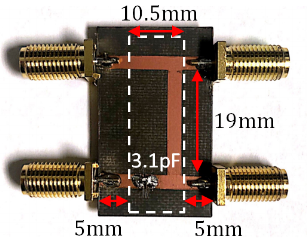}\label{fig:FB_UC}}}\subfigure[]{\textcolor{black}{\includegraphics[width=0.2\textwidth]{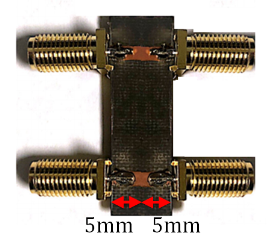}\label{fig:FB_CAL}}}
\par\end{centering}
\begin{centering}
\centering \subfigure[]{\textcolor{black}{\includegraphics[width=0.9\columnwidth]{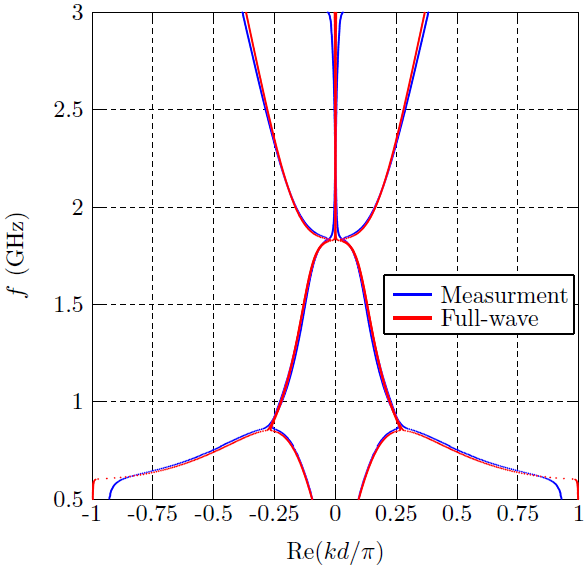}\label{fig:Dis_Mic-1-1-1}}}
\par\end{centering}
\textcolor{black}{\caption{\textcolor{black}{(a) Fabricated unit-cell for the CTL in Fig. \ref{fig:Micro_imp-1}
with 5mm extensions on both sides to be able to solder the SMA connectors.
(b) Fabricated microstrip extensions used for calibration, i.e., to
de-embed the effect of the extra extensions and SMA connectors from
(a). (c) Wavenumber dispersion versus frequency showing the existence
of the DBE around 1.85\ GHz, and two $2^{nd}$ order EPDs (i.e.,
regular band edges) around 0.86\ GHz. The measured result is in very
good agreement with that from full-wave simulations.}}
}
\end{figure}

\textcolor{black}{}
\textcolor{black}{To confirm the existence of EPDs in the periodic
CTL, we analyze a unit-cell and perform scattering (S)-parameter measurements
using a four-port Rohde \& Schwarz vector network analyzer (VNA) ZVA
67. Figure \ref{fig:FB_UC} shows the fabricated unit-cell with 5mm
extension on both sides to be able to solder the SMA connectors. The
measured scattering matrix is then transformed into a 4$\times$4
transfer matrix $\mathbf{\underline{T}}_{A}$. However this transfer
matrix, of the microstip in Fig. \ref{fig:FB_UC} that includes extensions,
is not the same as the transfer matrix of the one unit cell $\mathbf{\underline{T}}_{U}$,
however it is a cascaded version of it. The total transfer matrix
is $\mathbf{\underline{T}}_{A}=\mathbf{\underline{T}}_{R}\mathbf{\underline{T}}_{U}\mathbf{\underline{T}}_{L}$,
where $\mathbf{\underline{T}}_{R}$ and $\mathbf{\underline{T}}_{L}$
account for the extra lengths constituting the extensions at both
sides and the SMA connectors. In Fig. \ref{fig:FB_CAL} we show the
microstrip used in the two extensions, connected as a ``through'',
for calibration purposes \cite{apaydin2012experimental}. The transfer matrix of the two connected
extensions is $\mathbf{\underline{T}}_{B}=\mathbf{\underline{T}}_{R}\mathbf{\underline{T}}_{L}$.
Now a matrix that is proportional to the unit-cell transfer matrix
$\mathbf{\underline{T}}_{U}^{'}$ is obtained by de-embedding $\mathbf{\underline{T}}_{B}$
from $\mathbf{\underline{T}}_{A}$, i.e., $\mathbf{\underline{T}}_{U}^{'}=\mathbf{\underline{T}}_{A}\mathbf{\underline{T}}_{B}^{-1}=\mathbf{\underline{T}}_{R}\mathbf{\underline{T}}_{U}\mathbf{\underline{T}}_{R}^{-1}.$
It is important to point out that although $\mathbf{\underline{T}}_{U}$
and $\mathbf{\underline{T}}_{U}^{'}$ are not identical but they share
the same eigenvalues because $\mathbf{\underline{T}}_{U}^{'}$ is
just a transformed version of $\mathbf{\underline{T}}_{U}$ . Using
Floquet theory, following \cite{abdelshafy2018exceptional}, the dispersion
relation of the four modes is obtained as $e^{jkd}=\mathrm{eig}(\mathbf{\underline{T}}_{U})$
(i.e., the four eigenvalues of $\mathbf{\underline{T}}_{U}$) and
since $\mathbf{\underline{T}}_{U}$ and $\mathbf{\underline{T}}_{U}^{'}$
have identical eigenvalues, the dispersion is determined finally in
the form of $e^{jkd}=\mathrm{eig}(\mathbf{\underline{T}}_{U}^{'})=\mathrm{eig}(\mathbf{\underline{T}}_{A}\mathbf{\underline{T}}_{B}^{-1})$,
where $\mathbf{\underline{T}}_{A}$ and $\mathbf{\underline{T}}_{B}$
are the transfer matrices for the two four-port microstips in Fig.
\ref{fig:FB_UC} and Fig. \ref{fig:FB_CAL}, respectively. The wavenumber
dispersion diagram in Fig. \ref{fig:Dis_Mic-1-1-1} shows the four
coalescing complex wavenumbers (only the real parts are shown for
brevity, the imaginary parts is analogous to that in Fig. \ref{fig:case_1}).
In summary, the wavenumber dispersion diagram based on measurements
is in good agreement with the results based on the S-parameters calculated
via full-wave simulations based on the finite element method implemented
in CST Studio Suite. The dispersion shows several frequencies at which
EPD exists: a $4^{th}$ order EPD (the DBE) at $f\approx$1.85\ GHz
and two $2^{nd}$ order EPDs (the RBEs) at $f\approx$0.86\ GHz.
The perturbation due to ohmic, dielectric, and radiation losses seems
negligible because it does not destroy the occurrence of the EPDs. }

\begin{figure}
\begin{centering}
\centering \subfigure[]{\textcolor{black}{\includegraphics[width=0.88\columnwidth]{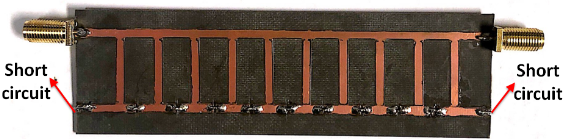}\label{fig:Dis_Mic-2-1}}}
\par\end{centering}
\begin{centering}
\centering\subfigure[]{\textcolor{black}{\includegraphics[width=0.9\columnwidth]{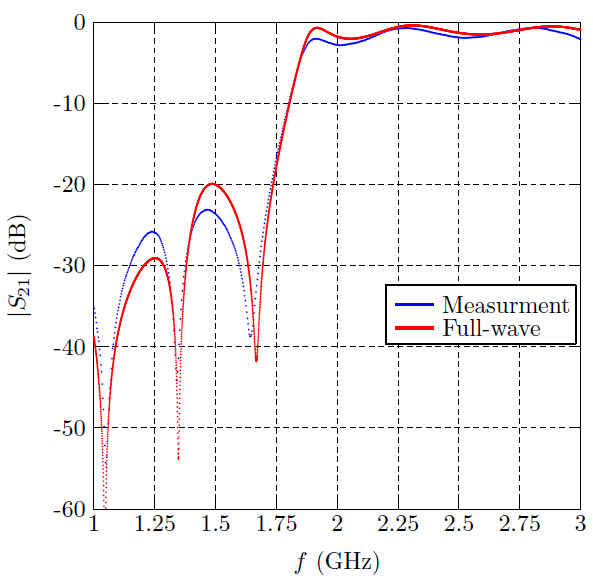}\label{fig:Dis_Mic-1-1-1-1}}}
\par\end{centering}
\textcolor{black}{\caption{\textcolor{black}{Measurements and simulations of the scattering parameter
$S_{21}$ for a nine-unit-cell CTL in (a). The result is consistent
with the DBE observation in the dispersion diagram at }\textit{\textcolor{black}{f}}\textcolor{black}{{}
=1.85\ GHz. The good agreement between full-wave simulations and
measurements shows that there is a DBE resonance associated with the
DBE.}}
}
\end{figure}

\textcolor{black}{}
\textcolor{black}{Figure \ref{fig:Dis_Mic-1-1-1} shows a nine-unit-cell
of the same DBE structure. The lower TL is connected to two short
circuits, similarly to the setup shown in Fig. \ref{fig:Q_1}. We
show in Fig. \ref{fig:Dis_Mic-1-1-1-1} the measurement and full-wave
simulation based on the finite element method, of the magnitude of
the scattering parameter $S_{21}$. These results show good agreement
between simulation and measurement. The results also demonstrate the
occurrence of the DBE resonance at 1.9\ GHz that is close to the
DBE frequency of 1.85\ GHz.}

\textcolor{black}{}

\section{\textcolor{black}{Conclusion}}

\textcolor{black}{We have shown the general conditions demonstrating
that a $4^{th}$ order EPD, namely a DBE, occurs at $k=0$ in two
}\textit{\textcolor{black}{uniform}}\textcolor{black}{{} lossless and
gainless CTLs when there is proper coupling between: (i) propagating
modes and evanescent modes, (ii) forward and backward propagating
modes, or (iii) four evanescent modes. We show that the resonance
frequency of a cavity made of a finite-length CTLs exhibiting a DBE
is very close to the DBE frequency, moreover, we show that the quality
factor increases with the fifth power of the cavity length (in the
lossless case) and such trend is robust to the occurrence of series
losses. Furthermore, we have shown that by using the CTL concept,
a }\textit{\textcolor{black}{regular band edge }}\textcolor{black}{can
be designed at non-vanishing wavenumbers. An example of CTLs supporting
the EPD wave phenomena discussed in this paper has been presented
using a metamaterial-based CTLs where the period to realize series
capacitances is sub-wavelength. We have provided the experimental
demonstration of the occurrence of the DBE in two uniform CTLs using
a metamaterial-like periodic CTL with subwavelength period, implemented
in microstrips. Possible applications exploiting the physics of the
DBE and the RBE are in high quality factors cavities \cite{nada2018giant},
radio frequency oscillators \cite{oshmarin2016oscillator} and distributed
oscillators \cite{abdelshafy2020distributed}, leaky wave antennas
\cite{othman2016theory}, filters, pulse compression \cite{F7}, sensors,
high power electron-beam devices \cite{abdelshafy2018electron}, and
lasers \cite{veysi2017theory}.}

\appendices{}

\textcolor{black}{\def\theequation{A\arabic{equation}}
\setcounter{equation}{0} }

\section{\textcolor{black}{General Solution of Wave Equation of Two Uniform
Coupled waveguides\label{sec:General-solution-of}}}

\textcolor{black}{Considering two coupled uniform TLs, the telegrapher's
equations that describe wave propagation are described by a first
order differential equation in (\ref{eq:telegraphic-1}). The general
solution of (\ref{eq:telegraphic-1}) with an initial condition $\mathbf{\Psi}_{zo}$
at $z=0$ is given by }

\textcolor{black}{
\begin{equation}
\mathbf{\Psi}(z)=\mathrm{exp}(-j\mathbf{\mathbf{\underline{M}}}z)\mathbf{\Psi}_{zo}.\label{eq:Appendix_Gen_Sol}
\end{equation}
}

\textcolor{black}{The matrix $\mathrm{exp}(-j\mathbf{\mathbf{\underline{M}}}z)$
is called transfer matrix. The system matrix $\mathbf{\mathbf{\underline{M}}}$
is diagonalizable when it has distinct eigenvectors, and the eigenvalues
$k_{1},$ $k_{2},$ $k_{3}$ and $k_{4},$ and the eigenvectors $\mathbf{\Psi}_{1},$
$\mathbf{\Psi}_{2},$ $\mathbf{\Psi}_{3}$ and $\mathbf{\Psi}_{4}$,
of $\mathbf{\mathbf{\underline{M}}}$ are determined by solving the
eigenvalue problem $\mathbf{\underline{M}}\mathbf{\Psi}=k\mathbf{\Psi}.$
The matrix $\mathrm{exp}(-j\mathbf{\mathbf{\underline{M}}}z)$ in
(\ref{eq:Appendix_Gen_Sol}) is generally determined by diagonalizing
the matrix $\mathbf{\underline{M}},$ however at EPDs where some of
the eigenvectors coalesce, the system matrix $\mathbf{\mathbf{\underline{M}}}$
can not be diagonalized and indeed the matrix $\mathbf{\mathbf{\underline{M}}}$
is similar to a matrix that contains at least a non trivial Jordan
block \cite{F3}, \cite{F5}.}

\subsection{\textcolor{black}{Diagonalizable System Matrix\label{subsec:Diagonalizable-System-matrix}}}

\textcolor{black}{When $\mathbf{\underline{M}}$ has distinct eigenvectors,
i.e., none of of the eigenmode coalesce, it can be diagonalized and
represented as}

\textcolor{black}{
\begin{equation}
\mathbf{\underline{M}=\underline{U}\ \underline{\Lambda}\ \underline{U}^{-1}},\label{eq:diag}
\end{equation}
where $\underline{\mathbf{U}}$ is the similarity transformation matrix
containing all the eigenvectors of $\mathbf{\underline{M}}$ as columns
and it is written in the form $\mathbf{\underline{U}=[\Psi_{\mathrm{1}}|\Psi_{\mathrm{2}}|\Psi_{\mathrm{3}}|\Psi_{\mathrm{4}}]}$,
whereas the matrix $\underline{\mathbf{\Lambda}}$ is a diagonal matrix
containing all the eigenvalues of $\mathbf{\underline{M}},$ viz.,
$\Lambda_{nn}=k_{n}$ for $n=1,2,3,4$. Since the eigenvectors of
the system are distinct, they form a complete set to represent any
state vector at any coordinate }\textit{\textcolor{black}{z}}\textcolor{black}{.
As a consequence, the initial condition $\mathbf{\Psi}_{zo}$ can
be represented as a linear decomposition of the eigenvectors (See
Ch.4 in \cite{Carl_Jordan}) as }

\textcolor{black}{
\begin{equation}
\mathbf{\Psi}_{zo}=a_{1}\mathbf{\Psi}_{1}+a_{2}\mathbf{\Psi}_{2}+a_{3}\mathbf{\Psi}_{3}+a_{4}\mathbf{\Psi}_{4}=\mathbf{\underline{U}\ a},\label{eq:Lin_Com}
\end{equation}
where $a_{n}$ are the the weights of each eigenvector, and the vector
$\mathbf{a}$ is written in the form $\mathbf{a}=[\begin{array}{cccc}
a_{1} & a_{2} & a_{3} & a_{4}\end{array}]^{\mathrm{T}}.$}

\textcolor{black}{Substituting (\ref{eq:diag}) and (\ref{eq:Lin_Com})
in (\ref{eq:Appendix_Gen_Sol}) yields}

\begin{equation}
\begin{array}{l}
\mathbf{\Psi}(z)=\mathbf{\underline{U}}\mathrm{exp}\left(-j\underline{\Lambda}z\right)\underline{\mathbf{U}}^{-1}\mathbf{\Psi}_{zo}.\ \\
\mathbf{\ \ \ \ \ \ \ =\underline{U}}\mathrm{exp}\left(-j\underline{\Lambda}z\right)\mathbf{a}\ \\
\ \ \ \ \ \ \ =[\mathbf{\Psi_{\mathrm{1}}}e^{-jk_{1}z}|\mathbf{\Psi}_{2}e^{-jk_{2}z}|\mathbf{\Psi}_{3}e^{-jk_{3}z}|\mathbf{\Psi}_{4}e^{-jk_{4}z}]\mathbf{a}\\
\begin{array}{c}
=a_{1}\mathbf{\Psi}_{1}e^{-jk_{1}z}+a_{2}\mathbf{\Psi}_{2}e^{-jk_{2}z}\ \ \ \ \ \ \ \ \ \ \ \ \\
\ \ \ \ \ \ \ \ \ \ \ \ \ \ \ \ \ \ \ \ \ \ \ \ \ +a_{3}\mathbf{\Psi}_{3}e^{-jk_{3}z}+a_{4}\mathbf{\Psi}_{4}e^{-jk_{4}z}.
\end{array}
\end{array}\label{eq:Mode_Sp}
\end{equation}

From (\ref{eq:Mode_Sp}), it is clear that the general solution of
the wave equation is decomposed of four eigenmodes, where each mode
separately is varying as $\Psi\propto e^{-jk_{n}z}$ .

\subsection{Non-Diagonalizable System Matrix with Fourth Order EPD}

At a fourth order EPD, the eigenvalues and the eigenvectors of $\underline{\mathbf{M}}$
coalesce, so $k_{{\color{black}n}}=k_{e}$ and $\mathbf{\Psi}_{n}=\mathbf{\Psi}_{e}$
for $n=1,2,3,4$, where $k_{e}$ and $\mathbf{\Psi}_{e}$ are the
degenerate eigenvalue and eigenvector, respectively. The system matrix
$\mathbf{\underline{M}}$ is not diagonalizable, whereas, the matrix
$\mathbf{U}$ constructed as described in the previous section at
any frequency near the EPD will be singular exactly at the EPD (as
a limit process). Hence the non-diagonizable $\mathbf{M}$ is similar
to a matrix in Jordan normal form (See Ch. 7 in \cite{Carl_Jordan})
as

\begin{equation}
\mathbf{\underline{M}=\underline{W}\ \left(\mathit{k_{e}}\underline{\mathbf{1}}+\underline{\mathbf{N}}\right)\ \underline{W}^{-1}}=\mathit{k_{e}}\underline{\mathbf{1}}+\underline{\mathbf{W}}\ \underline{\mathbf{N}}\ \underline{\mathbf{W}}^{-1},\label{eq:Non_diah}
\end{equation}
where $\underline{\mathbf{1}}$ is a $4\times4$ identity matrix,
and $\underline{\mathbf{N}}$ is a Nilpotent matrix, 

\begin{equation}
\underline{\mathbf{N}}=\left(\begin{array}{cccc}
0 & 1 & 0 & 0\\
0 & 0 & 1 & 0\\
0 & 0 & 0 & 1\\
0 & 0 & 0 & 0
\end{array}\right),
\end{equation}
and it follows the property

\begin{equation}
\begin{array}{cc}
\mathbf{\mathbf{N}}^{n}=\mathbf{0}, & \forall\ n\geq4\end{array}.\label{eq:power_Van_prop}
\end{equation}

The matrix $\underline{\mathbf{W}}$ contains the \textit{\textcolor{black}{generalized}}
eigenvectors of $\mathbf{\underline{M}}$ and is written in the form
$\mathbf{\underline{W}=[\Psi_{\mathrm{\mathit{e}}}|\Psi_{\mathrm{\mathit{e\mathrm{1}}}}|\Psi_{\mathrm{\mathit{e\mathrm{2}}}}|\Psi_{\mathrm{\mathit{e\mathrm{3}}}}]}$,
where 

\begin{equation}
\begin{array}{c}
\mathbf{\underline{M}\Psi_{\mathrm{\mathit{e}}}=0}\mathrm{,}\\
\mathbf{\underline{M}\Psi_{\mathrm{\mathit{e\mathrm{1}}}}=\Psi_{\mathrm{\mathit{e}}}},\\
\mathbf{\underline{M}\Psi_{\mathrm{\mathit{e\mathrm{2}}}}=\Psi_{\mathrm{\mathit{e\mathrm{1}}}}},\\
\mathbf{\underline{M}\Psi_{\mathrm{\mathit{e\mathrm{3}}}}=\Psi_{\mathrm{\mathit{\mathrm{e}\mathrm{2}}}}\mathrm{.}}
\end{array}\label{eq:Appendix_GEV}
\end{equation}

Substituting (\ref{eq:Non_diah}) in (\ref{eq:Appendix_Gen_Sol})
gives

\begin{equation}
\mathbf{\Psi}(z)=\mathrm{exp}\left(-jk_{e}z\underline{\mathbf{1}}-jz\underline{\mathbf{W}}\ \mathbf{\underline{N}}\ \underline{\mathbf{W}}^{\mathrm{-1}}\right)\mathrm{\Psi}_{zo}.\label{eq:GS_eq}
\end{equation}
Since the matrix $k_{e}\mathbf{\underline{\mathbf{1}}}$ and $\underline{\mathbf{W}}\ \mathbf{\underline{N}}\ \underline{\mathbf{W}}^{\mathrm{-1}}$
commute (See Ch. 10 in \cite{higham2008functions}), then (\ref{eq:GS_eq})
is simplified to 

\begin{equation}
\mathbf{\Psi}(z)=\mathrm{\mathit{e^{-jk_{e}z}}exp}\left(-jz\underline{\mathbf{W}}\ \mathbf{\underline{N}}\ \underline{\mathbf{W}}^{\mathrm{-1}}\right)\mathrm{\mathbf{\Psi}}_{zo}.\label{eq:eq_A_1}
\end{equation}
Using the Taylor series expansion of the exponential function (See
Ch. 10 in \cite{higham2008functions}) and using the fact that $\left(\underline{\mathbf{W}}\ \mathbf{\underline{N}}\ \underline{\mathbf{W}}^{\mathrm{-1}}\right)^{n}=\underline{\mathbf{W}}\ \mathbf{\underline{N}}^{n}\ \underline{\mathbf{W}}^{\mathrm{-1}}$
for any integer $n$, (\ref{eq:eq_A_1}) is expanded as 

\begin{equation}
\begin{array}{c}
\mathbf{\Psi}(z)=\mathit{e^{-jk_{e}z}}\sum\limits _{n=0}^{\infty}\dfrac{\underline{\mathbf{W}}\ \mathbf{\mathit{\mathrm{(}-jz}\underline{N})}^{n}\ \underline{\mathbf{W}}^{\mathrm{-1}}}{n!}\mathbf{\Psi}_{zo},\end{array}\label{eq:tr_mat-1-2}
\end{equation}
and making use of (\ref{eq:power_Van_prop}), (\ref{eq:tr_mat-1-2})
is reduced to 

\begin{equation}
\begin{array}{c}
\mathbf{\mathbf{\Psi}}(z)=\mathit{e^{-jk_{e}z}\underline{\mathbf{W}}}\left(\mathbf{\underline{\mathbf{1}}}-jz\mathbf{\underline{N}}-\dfrac{z^{2}\mathbf{\underline{N}}^{2}}{2}+\dfrac{jz^{3}\mathbf{\underline{N}}^{3}}{6}\right)\underline{\mathbf{W}}^{\mathrm{-1}}\mathbf{\Psi}_{zo}\end{array}\label{eq:tr_mat-1-2-2}
\end{equation}

At a fourth order EPD the state vector $\mathbf{\Psi}_{zo}$ at at
$z=0$ is represented as a series combination of the generalized eigenvectors
as 

\begin{equation}
\mathbf{\Psi}_{zo}=a_{e}\mathbf{\Psi}_{e}+a_{e1}\mathbf{\Psi}_{e1}+a_{e2}\mathbf{\Psi}_{e2}+a_{e3}\mathbf{\Psi}_{3}=\mathbf{\underline{W}\ a_{\mathit{e}}},\label{eq:Lin_Com-1}
\end{equation}
where $a_{en}$ are the the weights of the generalized eigenvectors,
and the vector $\mathbf{a_{\mathrm{\mathit{e}}}}$ is written in the
form $\mathbf{a_{\mathit{e}}}=[\begin{array}{cccc}
a_{e} & a_{e1} & a_{e3} & a_{e3}\end{array}]^{T}.$

Substituting (\ref{eq:Lin_Com-1}) into (\ref{eq:tr_mat-1-2}), the
general solution at fourth order EPD is obtained as 

\begin{equation}
\begin{array}{c}
\mathbf{\ \ \ \ \Psi_{\mathrm{e}}}(z)=\mathit{e^{-jk_{e}z}\underline{\mathbf{W}}}\left(\mathbf{\underline{\mathbf{1}}}-jz\mathbf{\underline{N}}-\dfrac{z^{2}\mathbf{\underline{N}}^{2}}{2}+\dfrac{jz^{3}\mathbf{\underline{N}}^{3}}{6}\right)\mathbf{a}_{e}\\
=\mathit{e^{-jk_{e}z}\underline{\mathbf{W}}}[\begin{array}{cccc}
a_{e}\ , & a_{e1}\ , & a_{e2}\ , & a_{e3}\end{array}]^{T}\\
\mathit{\ \ -jze^{-jk_{e}z}\underline{\mathbf{W}}}[\begin{array}{cccc}
a_{e1}\ , & a_{e2}\ , & a_{e3}\ , & 0\end{array}]^{T}\\
\ \ -\dfrac{z^{2}}{2}e^{-jk_{e}z}\underline{\mathbf{W}}[\begin{array}{cccc}
a_{e2}\ , & a_{e3}\ , & 0\ , & 0\end{array}]^{T}\\
-jz^{3}\mathit{e^{-jk_{e}z}\underline{\mathbf{W}}}[\begin{array}{cccc}
a_{e3}\ , & 0\ , & 0\ , & 0\end{array}]^{T}
\end{array}\label{eq:tr_mat-1-2-1}
\end{equation}

Simplifying (\ref{eq:tr_mat-1-2-1}), the general solution of (\ref{eq:telegraphic-1})
is cast in the form 

\begin{equation}
\begin{array}{l}
\mathbf{\Psi_{\mathrm{e}}}(z)=\ a_{e}\mathbf{\Psi}_{e}\mathit{e^{-jk_{e}z}}\\
\ \ \ \ \ \ \ \ \ \ +a_{e1}\left(\mathbf{\Psi}_{e1}-jz\mathbf{\Psi}_{e}\right)\mathit{e^{-jk_{e}z}}\\
\ \ \ \ \ \ \ \ \ \ +a_{e2}\left(\mathbf{\Psi}_{e2}-jz\mathbf{\Psi}_{e1}-\dfrac{z^{2}}{2}\mathbf{\Psi}_{e}\right)\mathit{e^{-jk_{e}z}}\\
\ \ \ \ \ \ \ \ \ \ +a_{e3}\Big(\mathbf{\Psi}_{e3}-jz\mathbf{\Psi}_{e2}-\dfrac{z^{2}}{2}\mathbf{\Psi}_{e1}+j\dfrac{z^{3}}{6}\mathbf{\Psi}_{e}\Big)\mathit{e^{-jk_{e}z}}.
\end{array}\label{eq:App_FEPD}
\end{equation}

From (\ref{eq:App_FEPD}), we conclude that only one mode preserve
the proportionality $\mathbf{\Psi}\propto e^{-jkz}$ at the fourth
order EPD, while the other three modes have algebraic growth with
$z$ as $\mathbf{\Psi}\propto\mathbf{P}(z)e^{-jkz}$, where $\mathbf{P}(z)$
is a polynomial vector funcion of maximum order of $3.$ 

For wavguides where the two equivalent CTLs are described by the per-unit
length parameters model as in Fig.\ref{fig:CCT_Model} the generalized
eigenvectors in (\ref{eq:App_FEPD}) are explicitly found to be

\begin{equation}
\begin{array}[t]{l}
\mathbf{\Psi}_{e}=[\begin{array}{cccc}
1\ , & -Y_{1}/Y_{2}\ , & 0\ , & 0\end{array}]^{T},\\
\mathbf{\Psi}_{e1}=[\begin{array}{cccc}
0\ , & 0\ , & j/Z_{1}\ , & jY_{2}/(Y_{1}Z_{1})\end{array}]^{T},\\
\mathbf{\Psi}_{e2}=[\begin{array}{cccc}
-(Y_{1}+Y_{2})/(Y_{1}^{2}Z_{1})\ , & 0\ , & 0\ , & 0\end{array}]^{T},\\
\mathbf{\Psi}_{e2}=[\begin{array}{cccc}
0\ , & 0\ , & -j(Y_{1}+Y_{2})/(Y_{1}^{2}Z_{1}^{2})\ , & 0\end{array}]^{T}.
\end{array}
\end{equation}

\section{Solution of Eigenvalue Problem for Uniform Coupled Waveguides\label{sec:Solution-of-Eigenvalue}}

Consider two \textit{uniform} CTLs described by generic per-unit-length
distributed parameters as shown in Fig. \ref{fig:CCT_Model}. In this
appendix we follow the derivation in \cite{yakovlev1998analysis}
to determine the wavenumbers of two \textit{uniform} CTLs. The wave
propagation in the structure is described by the first order differential
equations in (3). The wave equation describing wave propagation in
the two CTLs is obtained by taking the derivative of the first equation
in (3) with respect to $z$, and by inserting it into the second equation
of (3), leading to

\begin{equation}
\dfrac{d^{2}\mathbf{V}(z)}{dz^{2}}=\mathbf{\mathbf{\underline{\underline{Z}}}\mathrm{(}\omega\mathrm{)}\ }\mathbf{\underline{\underline{Y}}}\mathrm{(}\omega\mathrm{)}\mathbf{V}(z).\label{eq:telegraphic-2}
\end{equation}

The assumption of having propagating waves with function along $z$-direction
$\mathbf{V}(z)\propto e^{-jkz}$ makes the possible solutions of (\ref{eq:telegraphic-2})
cast in an eigenvalue problem form

\begin{equation}
k^{2}\mathbf{V}(z)=-\mathbf{\mathbf{\underline{\underline{Z}}}\mathrm{(}\omega\mathrm{)}\ }\mathbf{\underline{\underline{Y}}}\mathrm{(}\omega\mathrm{)}\mathbf{V}(z),\label{eq:Eig_V}
\end{equation}

Although the matrix $\mathbf{\underline{\underline{Z}}\ }\mathbf{\underline{\underline{Y}}}$
is a $2\times2$ matrix the eigenvalues $k$ obtained from (\ref{eq:Eig_V})
are four and they are identical to those obtained from (\ref{eq:Eig_Prop}).
From (\ref{eq:Eig_V}) it is clear that eigenmodes satisfy the $\pm k$
symmetry. They represent two waves that can propagate or attenuate
along each positive and negative $z$-directions, i.e., four modes.
The characteristic equation of the eigenvalue problem in (\ref{eq:Eig_V}),
which represents the dispersion relation of the structure, can be
written in their simplest form as

\begin{equation}
k^{4}+Tk^{2}+D=0.
\end{equation}
Therefore, the four roots of the above equation, wavenumbers, can
finally we written as in (\ref{eq:dis}).

The eigenvectors of (\ref{eq:Eig_V}) may be written in their simplest
form as

\begin{equation}
\mathbf{V}_{n}=\psi_{0}\left(\begin{array}{c}
Z_{1}\left(k_{n}^{2}+Z_{2}(Y_{2}+Y_{c})\right)\\
Z_{1}Z_{2}Y_{c}
\end{array}\right),\label{eq:Eig_Ve-2}
\end{equation}

where $\psi_{0}$ is arbitrary constant and it has a unit of $\mathrm{Am}^{3}$.
It is important to point about that the eigenvectors representing
voltages propagating along the negative $z$-direction are identical
to those in positive $z$-direction due to the fact that the structure
is reciprocal, however, their corresponding current vectors are not
identical to each others, and indeed they have a sign difference,
and are determined from (3) as

\begin{equation}
\mathbf{I}_{n}=jk_{n}\mathbf{\underline{\underline{Z}}}^{-1}\mathbf{V}_{n}=\psi_{0}\left(\begin{array}{c}
jk_{n}\left(k_{n}^{2}+Z_{2}(Y_{2}+Y_{c})\right)\\
jZ_{1}k_{n}Y_{c}
\end{array}\right).\label{eq:Eig_Ve-2-1}
\end{equation}

Combining the eigenvectors in (\ref{eq:Eig_Ve-2}) and (\ref{eq:Eig_Ve-2-1}),
the four eigenvectors of the eigenvalue problem in (\ref{eq:Eig_Prop})
are found in their simplest form as in (\ref{eq:Eig_Ve})

\bibliographystyle{ieeetr}
\bibliography{EPD_General_Cond_Ref}

\begin{thebibliography}{10}

\bibitem{F3}
A.~Figotin and I.~Vitebskiy, ``Frozen light in photonic crystals with
  degenerate band edge,'' {\em Physical Review E}, vol.~74, no.~6, p.~066613,
  2006.

\bibitem{2017theory}
M.~Y. Nada, M.~A. Othman, and F.~Capolino, ``Theory of coupled resonator
  optical waveguides exhibiting high-order exceptional points of degeneracy,''
  {\em Physical Review B}, vol.~96, no.~18, p.~184304, 2017.

\bibitem{othman2016giant}
M.~A. Othman, F.~Yazdi, A.~Figotin, and F.~Capolino, ``Giant gain enhancement
  in photonic crystals with a degenerate band edge,'' {\em Physical Review B},
  vol.~93, no.~2, p.~024301, 2016.

\bibitem{F2}
A.~Figotin and I.~Vitebskiy, ``Gigantic transmission band-edge resonance in
  periodic stacks of anisotropic layers,'' {\em Physical Review E}, vol.~72,
  no.~3, p.~036619, 2005.

\bibitem{nada2018giant}
M.~Y. Nada, M.~A. Othman, O.~Boyraz, and F.~Capolino, ``Giant resonance and
  anomalous quality factor scaling in degenerate band edge coupled resonator
  optical waveguides,'' {\em Journal of Lightwave Technology}, vol.~36, no.~14,
  pp.~3030--3039, 2018.

\bibitem{figotin2003oblique}
A.~Figotin and I.~Vitebskiy, ``Oblique frozen modes in periodic layered
  media,'' {\em Physical Review E}, vol.~68, no.~3, p.~036609, 2003.

\bibitem{Hamid_EPD_LTV}
H.~Kazemi, M.~Y. Nada, T.~Mealy, A.~F. Abdelshafy, and F.~Capolino,
  ``Exceptional points of degeneracy induced by linear time-periodic
  variation,'' {\em Physical Review Applied}, vol.~11, no.~1, p.~014007, 2019.

\bibitem{el2007theory}
R.~El-Ganainy, K.~Makris, D.~Christodoulides, and Z.~H. Musslimani, ``Theory of
  coupled optical$\text{ PT-}$symmetric structures,'' {\em Optics Letters},
  vol.~32, no.~17, pp.~2632--2634, 2007.

\bibitem{othman2016theory}
M.~A. Othman and F.~Capolino, ``Theory of exceptional points of degeneracy in
  uniform coupled waveguides and balance of gain and loss,'' {\em IEEE
  Transactions on Antennas and Propagation}, vol.~65, no.~10, pp.~5289--5302,
  2017.

\bibitem{F4}
G.~Mumcu, K.~Sertel, and J.~L. Volakis, ``Miniature antenna using printed
  coupled lines emulating degenerate band edge crystals,'' {\em IEEE
  Transactions on Antennas and Propagation}, vol.~57, no.~6, pp.~1618--1624,
  2009.

\bibitem{F5}
M.~A. Othman, F.~Yazdi, A.~Figotin, and F.~Capolino, ``Giant gain enhancement
  in photonic crystals with a degenerate band edge,'' {\em Physical Review B},
  vol.~93, no.~2, p.~024301, 2016.

\bibitem{othman2015demonstration}
M.~A. Othman and F.~Capolino, ``Demonstration of a degenerate band edge in
  periodically-loaded circular waveguides,'' {\em IEEE Microwave and Wireless
  Components Letters}, vol.~25, no.~11, pp.~700--702, 2015.

\bibitem{zheng2019design}
T.~Zheng, M.~Casaletti, A.~F. Abdelshafy, F.~Capolino, Z.~Ren, and G.~Valerio,
  ``Design of substrate integrated waveguides supporting degenerate band-edge
  resonances,'' in {\em 2019 13th European Conference on Antennas and
  Propagation (EuCAP)}, Krakow, Poland, 2019, pp. 1-3.

\bibitem{F6}
C.~{Locker}, K.~{Sertel}, and J.~L. {Volakis}, ``Emulation of propagation in
  layered anisotropic media with equivalent coupled microstrip lines,'' {\em
  IEEE Microwave and Wireless Components Letters}, vol.~16, no.~12,
  pp.~642--644, 2006.

\bibitem{F7}
V.~A. Tamma, A.~Figotin, and F.~Capolino, ``Concept for pulse compression
  device using structured spatial energy distribution,'' {\em IEEE Transactions
  on Microwave Theory and Techniques}, vol.~64, no.~3, pp.~742--755, 2016.

\bibitem{sloan2017theory}
J.~T. Sloan, M.~A. Othman, and F.~Capolino, ``Theory of double ladder lumped
  circuits with degenerate band edge,'' {\em IEEE Transactions on Circuits and
  Systems I: Regular Papers}, vol.~65, no.~1, pp.~3--13, 2017.

\bibitem{burr2013degenerate}
J.~R. Burr, N.~Gutman, C.~M. de~Sterke, I.~Vitebskiy, and R.~M. Reano,
  ``Degenerate band edge resonances in coupled periodic silicon optical
  waveguides,'' {\em Optics Express}, vol.~21, no.~7, pp.~8736--8745, 2013.

\bibitem{othman2017experimental}
M.~A. Othman, X.~Pan, G.~Atmatzakis, C.~G. Christodoulou, and F.~Capolino,
  ``Experimental demonstration of degenerate band edge in metallic periodically
  loaded circular waveguide,'' {\em IEEE Transactions on Microwave Theory and
  Techniques}, vol.~65, no.~11, pp.~4037--4045, 2017.

\bibitem{abdelshafy2018exceptional}
A.~F. {Abdelshafy}, M.~A.~K. {Othman}, D.~{Oshmarin}, A.~T. {Almutawa}, and
  F.~{Capolino}, ``Exceptional points of degeneracy in periodic coupled
  waveguides and the interplay of gain and radiation loss: Theoretical and
  experimental demonstration,'' {\em IEEE Transactions on Antennas and
  Propagation}, vol.~67, no.~11, pp.~6909--6923, 2019.

\bibitem{yazdi2017new}
F.~Yazdi, M.~A. Othman, M.~Veysi, A.~Figotin, and F.~Capolino, ``A new
  amplification regime for traveling wave tubes with third-order modal
  degeneracy,'' {\em IEEE Transactions on Plasma Science}, vol.~46, no.~1,
  pp.~43--56, 2017.

\bibitem{abdelshafy2018electron}
A.~F. Abdelshafy, M.~A. Othman, F.~Yazdi, M.~Veysi, A.~Figotin, and
  F.~Capolino, ``Electron-beam-driven devices with synchronous multiple
  degenerate eigenmodes,'' {\em IEEE Transactions on Plasma Science}, vol.~46,
  no.~8, pp.~3126--3138, 2018.

\bibitem{oshmarin2016oscillator}
D.~Oshmarin, F.~Yazdi, M.~A. Othman, J.~Sloan, M.~Radfar, M.~M. Green, and
  F.~Capolino, ``New oscillator concept based on band edge degeneracy in lumped
  double-ladder circuits,'' {\em IET Circuits, Devices \& Systems}, vol.~13,
  no.~7, pp.~950--957, 2019.

\bibitem{veysi2017theory}
M.~Veysi, M.~A. Othman, A.~Figotin, and F.~Capolino, ``Degenerate band edge
  laser,'' {\em Physical Review B}, vol.~97, no.~19, p.~195107, 2018.

\bibitem{mealy2019degeneracy}
T.~Mealy, A.~F. Abdelshafy, and F.~Capolino, ``The degeneracy of the dominant
  mode in rectangular waveguide,'' in {\em 2019 United States National
  Committee of URSI National Radio Science Meeting (USNC-URSI NRSM)}, Boulder,
  CO, USA, 2019, pp. 1-2.

\bibitem{bender1998real}
C.~M. Bender and S.~Boettcher, ``Real spectra in non-$\text{H}$ermitian
  $\text{H}$amiltonians having $\text{PT}$ symmetry,'' {\em Physical Review
  Letters}, vol.~80, no.~24, p.~5243, 1998.

\bibitem{paul2008analysis}
C.~R. Paul, {\em Analysis of multiconductor transmission lines}.
\newblock Hoboken, NJ, USA: Wiley, 2008.

\bibitem{yakovlev1998analysis}
G.~W. {Hanson}, A.~B. {Yakovlev}, M.~A.~K. {Othman}, and F.~{Capolino},
  ``Exceptional points of degeneracy and branch points for coupled transmission
  lines--linear-algebra and bifurcation theory perspectives,'' {\em IEEE
  Transactions on Antennas and Propagation}, vol.~67, no.~2, pp.~1025--1034,
  2019.

\bibitem{5340521}
G.~{Mumcu}, K.~{Sertel}, and J.~L. {Volakis}, ``Lumped circuit models for
  degenerate band edge and magnetic photonic crystals,'' {\em IEEE Microwave
  and Wireless Components Letters}, vol.~20, no.~1, pp.~4--6, 2010.

\bibitem{Wave_Scenarios}
I.~A. Eshrah, A.~A. Kishk, A.~B. Yakovlev, and A.~W. Glisson, ``Evanescent
  rectangular waveguide with corrugated walls: A composite right/left-handed
  metaguide,'' in {\em Microwave Symposium Digest, 2005 IEEE MTT-S
  International}, Long Beach, CA, 2005, p. 4.

\bibitem{noh2010giant}
H.~Noh, J.~Yang, I.~Vitebskiy, A.~Figotin, and H.~Cao, ``Giant resonances near
  the split band edges of two-dimensional photonic crystals,'' {\em Physical
  Review A}, vol.~82, no.~1, p.~013801, 2010.

\bibitem{Quality_Ref}
B.~Razavi, ``A study of phase noise in cmos oscillators,'' {\em IEEE journal of
  Solid-State circuits}, vol.~31, no.~3, pp.~331--343, 1996.

\bibitem{abdelshafy2020distributed}
A.~F. Abdelshafy, D.~Oshmarin, M.~A. Othman, M.~M. Green, and F.~Capolino,
  ``Distributed degenerate band edge oscillator,'' {\em arXiv preprint
  arXiv:2002.00857}, 2020.

\bibitem{apaydin2012experimental}
N.~Apaydin, L.~Zhang, K.~Sertel, and J.~L. Volakis, ``Experimental validation
  of frozen modes guided on printed coupled transmission lines,'' {\em IEEE
  transactions on microwave theory and techniques}, vol.~60, no.~6,
  pp.~1513--1519, 2012.

\bibitem{Carl_Jordan}
C.~D. Meyer, {\em Matrix analysis and applied linear algebra}.
\newblock Philadelphia, PA, USA: SIAM, 2001.

\bibitem{higham2008functions}
N.~Higham, {\em Functions of matrices : theory and computation}.
\newblock Philadelphia, PA, USA: SIAM, 2008.

\end{thebibliography}

\end{document}